\documentclass[
]{ceurart}

\usepackage{listings}
\usepackage{subcaption}
\usepackage[section]{placeins}

\usepackage{todonotes}
\usepackage{xspace}

\newcommand{\bve}{BVE\(^+\)\xspace}

\usepackage[binary-units=true,group-four-digits=true,detect-all]{siunitx}

\begin{document}

\copyrightyear{2023}
\copyrightclause{Copyright for this paper by its authors.
  Use permitted under Creative Commons License Attribution 4.0
  International (CC BY 4.0).}

\conference{14th Pragmatics of SAT international workshop, a workshop of the 26th International Conference on Theory and Applications of Satisfiability Testing, July 04, 2023, Alghero. Italy} 

\title{Life span of SAT techniques}


\author[1,2]{Mathias Fleury}[%
orcid=0000--0002--1705--3083,
email=fleury@cs.uni-freiburg.de,
]
\cormark[1]

\address[1]{University of Freiburg, Freiburg im Breisgau, Germany}
\address[2]{JKU Linz, Linz, Austria}

\author[3]{Daniela Kaufmann}[%
orcid=0000--0002--5645--0292,
email=daniela.kaufmann@tuwien.ac.at,
]
\address[3]{TU Wien, Vienna, Austria}

\cortext[1]{Corresponding author.}

\begin{abstract}
  In this paper we take 4 different features of the SAT solver
  CaDiCaL, blocked clause elimination, vivification, on-the-fly self
  subsumption, and increasing the bound of variable elimination  
  over the SAT Competitions benchmarks between 2009 and 2022.  We
  study these features by both activating them one-by-one and
  deactivating them one-by-one.  
  We have three hypothesis regarding the experiments: (i) disabling features is always harmful; (ii) the life span of the techniques is limited; and (iii) features simulate each other. Our experiments cannot confirm any of the hypothesis.
  \end{abstract}


\maketitle
\section{Introduction}

Over the years, numerous methods have been suggested to solve more instances on paper or as submissions to the SAT Competition.
All of them have shown some improvements when proposed. But most of them
never became mainstream.
Moreover, many techniques are now no longer used in state-of-the-art SAT
solvers -- they fell out of fashion because SAT implementers
considered them detrimental for the SAT Competition.

\paragraph{Our Study.} In this paper we attempt a systematic study: We take only four
techniques (Section~\ref{sec:techniques}), two that are part of many
recent SAT solvers (\emph{vivification} and 
\emph{variable elimination with bound
increase, \bve}~\cite{DBLP:journals/jacm/DavisP60}), one that is part of some but not all (\emph{on-the-fly self
subsumption, OTFS}~\cite{otfs1,otfs2}), and one that is not part anymore (\emph{block clause elimination, BCE}~\cite{bce,bcerecons}).

To test the techniques, we use the SAT solver CaDiCaL~\cite{BiereFazekasFleuryHeisinger-SAT-Competition-2020-solvers}. CaDiCaL already supports three of the considered techniques and we only had to implement
OTFS. Yet, it contains many inprocessing techniques that are not entirely
independent of the techniques that we have removed (Section~\ref{sec:cadical}).

We work in two different directions: (i) we deactivate all 4
techniques and activate only one at a time; and (ii) we only deactivate one technique at a time. For the ablation approach~(ii) we use two starting points; the first being the default version of CaDiCaL (OTFS, variable elimination with bound increase, vivification), the second being the default version of CaDiCaL with BCE.

The ablation approach (ii) is
the one used in most papers to show how useful a new technique is. We
run the various versions over all the problems from the SAT
Competition between 2009 and 2022 (Section~\ref{sec:results}).

\paragraph{Organization.} The remainder of the paper is organized as follows. In Section~\ref{sec:techniques} we briefly discuss the four considered techniques in our study. Section~\ref{sec:cadical} gives an introduction to the solver CaDiCaL. We provide our expectations of this study in Section~\ref{sec:expectations}, before we present our results in Section~\ref{sec:results} and conclude with our interpretation in Section~\ref{sec:conclusion}.

This paper is a minor extension of our (unpublished) talk at the
POS'23 workshop. We provide the data for further analysis as Zenodo
artifact~\cite{fleury_2024_10608480}.

  
\section{Techniques under investigation}\label{sec:techniques}

In this Section we provide an introduction to the considered techniques OTFS, BCE, \bve,  and vivification.

\paragraph{On-the-fly Self
Subsumption.}
On-the-fly self-subsumption is a technique dating back to 2009~\cite{otfs1,otfs2}
that strengthens clauses during the conflict analysis. This is similar
to subsumption, but it can be done cheaply during the usual conflict analysis.

The core idea is that CDCL is working on the conflicting clause \(L\lor C\)
and resolves it with the reason of \(L\), the clause \(\neg L \lor D\).
Usually both clauses are simply resolved together to get \(C\lor D\).
With OTFS we also check wether the clause \(C\lor D\) is shorter than
\(L\lor C\). If this is the case, the solver removes the literal \(L\) from the
clause \(L\lor C\). In some cases, the new clause \(C\lor D\) is now a
missed propagation without learning a new clause.

Combining this technique with usual SAT solvers does not break
invariants, unless the SAT solver contains chronological
backtracking~\cite{chrono1,chrono2}. In the latter case, the conflict
analysis can produce a clause that is conflicting on a lower level and
requires backtracking on this lower level to continue the conflict analysis.
Without it, CDCL eagerly propagates so the clause \(L\lor C\) contains at least two literals on current level.
With it, the clause may contain only \emph{one} literal on current level. If this literal
is removed, the clause is a ``missed'' conflict.
Besides that, OTFS and
chronological backtracking break similar invariants -- both can find
missed propagations.

This technique is implemented in some recent SAT solvers including
Kissat, but not in CaDiCaL nor in Minisat/Glucose/MapleSAT (to the best of our knowledge).
The folklore knowledge seems to be that it is a beautiful and tempting idea, but
with only a limited effect.

\paragraph{Blocked clause elimination.}
Blocked clause elimination is a technique to remove clauses from
the problem.  Removing these clauses does not change the satisfiability of the
problem -- but models must be reconstructed to be models of the
initial set of clauses~\cite{bce,bcerecons}.

A clause \(L\lor C\) is \emph{blocked} with respect to \(L\) whenever all resolvents on
\(L\) are tautological. In those cases, the clause \(L\lor C\) can
simply be removed (and the clause set remains SAT/UNSAT).

This approach is not activated by default in CaDiCaL and is not implemented in Minisat/Glucose/MapleSAT.

\paragraph{Variable elimination with Bound Increase.}
Variable elimination is the base of the venerable DP
algorithm~\cite{DBLP:journals/jacm/DavisP60}. It is a decision procedure but is restricted in
most SAT solvers to add only few clauses (usually under the name bounded variable elimination
BVE). Since Minisat in 2006, it is used as preprocessor to
heuristically simplify the problems before running CDCL.  It can also
be used as an inprocessing technique during the
search~\cite{BiereJarvisaloKiesl-SAT-Handbook-2021}.

Elimination consists in resolving all clauses on a given literal --
excluding tautologies.  The standard restriction is that after
eliminating a literal the number of clauses is at most the same --
even if the clauses become longer. This superseeds pure literal
deletion.

Since 2015, the abcdSAT SAT solver lifted the limitation:
elimination can (slowly) increase the number of clauses. In CaDiCaL,
the limitation is slowly increased to 16: elimination of a literal can at most generate 16
clauses more than were initially present. We refer to this technique
as \bve to distinguish it from the default BVE.

This approach is implemented in CaDiCaL, Kissat, and recent MapleSAT solvers
with different approaches: CaDiCaL slowly increases the limit to 16,
while MapleSAT directly allows for 20 new clauses.

\paragraph{Vivification.}
Vivification~\cite{DBLP:journals/ai/LiXLMLL20} is a technique that
reuses a SAT solver core procedure: propagation and conflict
analysis. It takes a clause, negates the literals, and decides them
one-by-one. Between the decision, the usual propagation loop is used
(ignoring the clause to vivify).
As in the standard case, a new clause can be learned by the usual conflict analysis.
The peculiarity in that case is that the result
can be used to strengthen the original clause -- in
particular propagated literals can be removed to strengthen the clause.

This approach is implemented in CaDiCaL, Kissat, and recent MapleSAT
solvers with different approaches. Kissat vivifies parts of the
clauses (depending on the years, redundant only or irredundant and
redundant clauses). MapleSAT also vivifies clauses (core and tier 2)
with a different scheduling. CaDiCaL vivifies redundant and irredundant
clauses separately (but does not distinguish between tier 2 and tier 3 clauses).


\section{CaDiCaL}\label{sec:cadical}

For our experiments we use the SAT solver CaDiCaL~\cite{BiereFazekasFleuryHeisinger-SAT-Competition-2020-solvers}. While initially
designed to be a \emph{radically simplified} CDCL SAT solver, it
offers many features and is very readable. It is since 2021 the base of the
hack-track of the SAT Competition.

For our experiments we keep the solver CaDiCaL with all its features
and only deactivate some or activate BCE (that is off by default).
Here is a short list of the inprocessing features that are similar to some of the features we test:

\begin{description}
\item [Forward and backward subsumption-resolution] resolves two clauses together
  in an attempt to shorten one of them. Vivification can be seen as a
  generalization of this technique: it makes it possible to remove
  several literals at once instead of a single one. However, to simulate vivification, resolution must be allowed to
  do several resolution steps at once, even if the intermediate
  steps are longer.
\item [Failed literal probing] 
makes binary clauses out of longer clauses or adds new
  binary clauses. It can partially emulate vivification, when only one
  decision is allowed.
\item[Exhaustive ternary resolution] resolves all ternary clauses together
  that produce a new binary or a new ternary clause. It makes
  subsumption-resolution stronger, but still weaker than vivification.
\end{description}

For the sake of the experiment we added OTFS
to CaDiCaL. The implementation was easy thanks to the built-in checker
and the built-in model-based tester Mobical. We expect to merge our development in the next CaDiCaL release.



      
   




In the following, we consider three configurations: 
\begin{itemize}
\item The \emph{base}
configuration that deactivates all four considered features, yet leaving all 
other features of CaDiCaL with the same defaults.
\item The \emph{default}
configuration; using exactly the options given by default with OTFS,
vivification, and \bve, but not BCE.
\item The \emph{everything}
configuration; using exactly the options given by default with OTFS,
vivification, \bve, and BCE.
\end{itemize}

The choice to include OTFS in the default configuration comes from the
fact that we know from previous experience that it has little effect.

\section{Expectations}
\label{sec:expectations}

We had the following hypotheses what we are going to see in the experiments.

\begin{itemize}
\item \textbf{Hypothesis 1 - Disabling features decreases number of solved instances.} From default we were expecting to see drops when one feature is turned off, as this would show that all of them are needed.

\item \textbf{Hypothesis 2 - Life span of techniques is limited.} For base, we were expecting that adding a particular feature shows an increase in the solved instances in the same time area when the particular feature was invented, which reduces to zero after 2-3 years. The reason why we expected this behaviour was that the inventor of the feature proved the feature to be useful in year X. Then every other SAT solver integrates the same technique in year X+1. Hence, every solver supports the technique and will be able to perform well on benchmarks where said feature is important. So in years X+2 new benchmarks do not address this feature anymore.  

\item \textbf{Hypothesis 3 - Features simulate each other.} 
We propose to see that two techniques perform similar, as this shows that one technique simulates another feature. If we do not see a dependency between the techniques, this means that all techniques stand on their own and there is no correlation in the results.
\end{itemize}

When entering the experiments we were not sure if deactivating a feature from default would also show some benchmark bias or not. 

One limitation is that CaDiCaL was trained by Biere on the benchmarks from the SAT Competition 2016 to 2018, meaning that we can
expect the default options to work well together.

\section{Results}\label{sec:results}

In this section we present our experimental results. As discussed in previous Sections, we conduct three experimental settings:
\begin{itemize}
  \item \emph{Experiment ``base''.} In our first setting we show the effects the individual techniques have, when we compare them with the \emph{base} version of CaDiCaL, where none of the four techniques is turned on. The tested versions are ``base'', ``base+BCE'', ``base+vivify'', ``base+OTFS'', and ``base+\bve''. As explained above (Section~\ref{sec:cadical}), there are many features still remaining.
\item \emph{Experiment ``default''.} We compare the effect of each feature compared to the \emph{default} setting of CaDiCaL. We again want to highlight that the default strategy of CaDiCaL consists of vivication, \bve, and OTFS being enabled, whereas BCE is turned of. Hence, in this experimental setting we compare the default strategy of CaDiCaL to the versions where vivification, \bve, or OTFS are disabled individually. Hence we consider the versions ``default'', ``default-vivify'', ``default-OTFS'', and ``default-\bve'' (read the ``-'' as minus).
\item \emph{Experiment ``everything''.} We compare the effect of each feature compared to the \emph{everything-on} setting of CaDiCaL (= default + BCE). In this setting we compare CaDiCaL to the versions where either BCE, vivification, \bve, or OTFS are disabled individually. Hence we compare the versions ``everything'', ``everything-BCE (=default)'', ``everything-vivify'', ``everything-OTFS'', and ``everything-\bve''.
\end{itemize}

In our  experiments  we use  an  Intel Xeon  E5-2620  v4  CPU  at \SI{2.10}{\giga\hertz} (with turbo-mode disabled) with a memory limit of \SI{128}{\giga\byte}. As benchmarks we use the benchmarks of the SAT competition of the years 2009-2022. 
We use the same setting as in the recent annual SAT competition and provide a time limit of \SI{5000}{\second} for each benchmark to be solved. 

The summary of our results can be seen in Table~\ref{tab:all-results}.
By ``\#Benchmarks'' we denote the total number of benchmarks in the SAT competition of the corresponding year.
The reported numbers present the number of instances CaDiCaL is able to solve within the given time limit. 

\tabcolsep1mm
\begin{table}[ht!]
  \centering
  \begin{tabular}{lrrrrrrrrrrrrrrr}
\toprule
 & 2009 & 2010 & 2011 & 2012 & 2013 & 2014 & 2015 & 2016 & 2017 & 2018 & 2019 & 2020 & 2021 & 2022 & 2023 \\
\midrule
\#Benchmarks & 292 & 101 & 300 & 600 & 293 & 300 & 300 & 300 & 350 & 400 & 400 & 400 & 400 & 400 & 400 \\
evthg & 227 & 86 & 237 & 559 & 238 & 244 & 272 & 163 & 231 & 272 & 238 & 235 & 258 & 260 & 239 \\
evthg-BCE (default) & 229 & 88 & 235 & 558 & 235 & 243 & 268 & 166 & 235 & 271 & 247 & 229 & 256 & 267 & 237 \\
evthg-BVE+ & 227 & 87 & 235 & 559 & 234 & 246 & 267 & 163 & 230 & 270 & 241 & 237 & 256 & 261 & 244 \\
evthg-OTFS & 229 & 85 & 240 & 559 & 237 & 243 & 267 & 168 & 229 & 272 & 235 & 234 & 262 & 258 & 236 \\
evthg-vivify & 228 & 89 & 241 & 559 & 236 & 246 & 268 & 158 & 230 & 271 & 235 & 233 & 261 & 264 & 240 \\
default & 229 & 88 & 235 & 558 & 235 & 243 & 268 & 166 & 235 & 271 & 247 & 229 & 256 & 267 & 237 \\
default-vivify & 230 & 88 & 238 & 560 & 237 & 244 & 271 & 161 & 239 & 271 & 234 & 236 & 262 & 263 & 242 \\
default-OTFS & 229 & 87 & 240 & 561 & 235 & 242 & 269 & 169 & 236 & 269 & 235 & 233 & 259 & 260 & 236 \\
default-BVE+ & 228 & 88 & 235 & 558 & 238 & 244 & 268 & 164 & 231 & 268 & 243 & 236 & 259 & 258 & 239 \\
default-truephase & 218 & 84 & 214 & 531 & 227 & 225 & 255 & 150 & 198 & 205 & 218 & 173 & 232 & 231 & 179 \\
default-falsephase & 208 & 81 & 208 & 498 & 208 & 199 & 243 & 149 & 200 & 211 & 212 & 185 & 232 & 232 & 185 \\
base & 230 & 87 & 239 & 558 & 228 & 236 & 264 & 165 & 234 & 270 & 243 & 230 & 263 & 263 & 243 \\
base+vivify & 228 & 86 & 237 & 555 & 230 & 240 & 268 & 165 & 232 & 264 & 244 & 240 & 259 & 262 & 231 \\
base+OTFS & 231 & 89 & 241 & 562 & 231 & 242 & 265 & 161 & 238 & 274 & 239 & 235 & 262 & 265 & 243 \\
base+BVE+ & 232 & 87 & 237 & 559 & 237 & 243 & 266 & 163 & 233 & 270 & 240 & 231 & 254 & 260 & 242 \\
base+BCE & 227 & 87 & 238 & 555 & 228 & 238 & 260 & 166 & 232 & 268 & 239 & 236 & 261 & 259 & 244 \\
\bottomrule
\end{tabular}

  \caption{Raw performance of the different configurations over the years.}
  \label{tab:all-results}
\end{table}

\begin{figure}[ht!]
  \centering
  \begin{subfigure}{.5\textwidth}
    \centering
    \includegraphics[height=.7\linewidth]{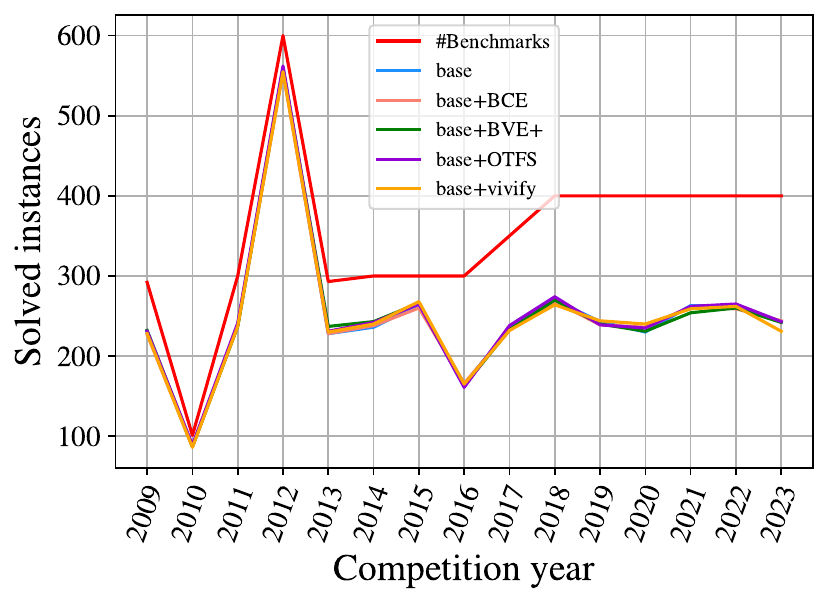}
    \caption{Comparison with base}
    \label{fig:base-all}
  \end{subfigure}
  \begin{subfigure}{.5\textwidth}
    \centering
    \includegraphics[height=.7\linewidth]{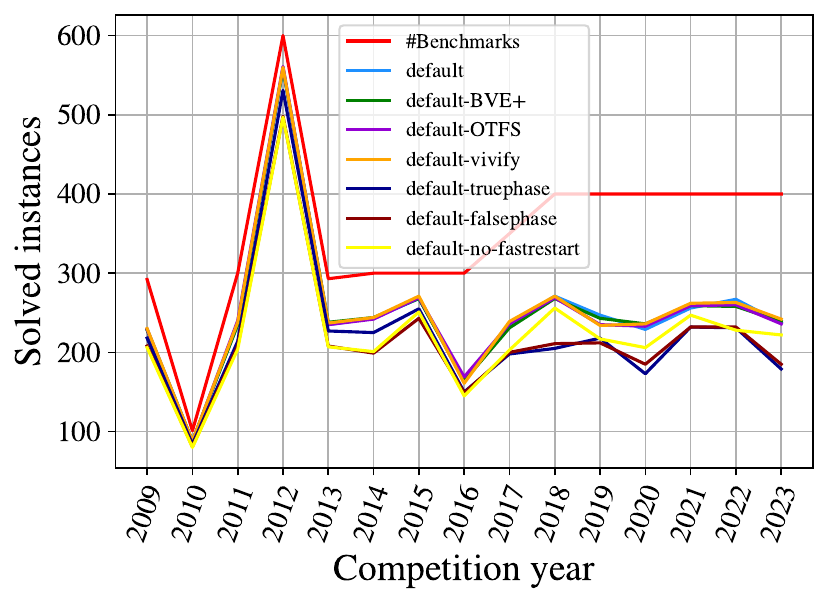}
    \caption{Comparison with default}
    \label{fig:def-all} 
  \end{subfigure}%
  \begin{subfigure}{.5\textwidth}
    \centering
    \includegraphics[height=.7\linewidth]{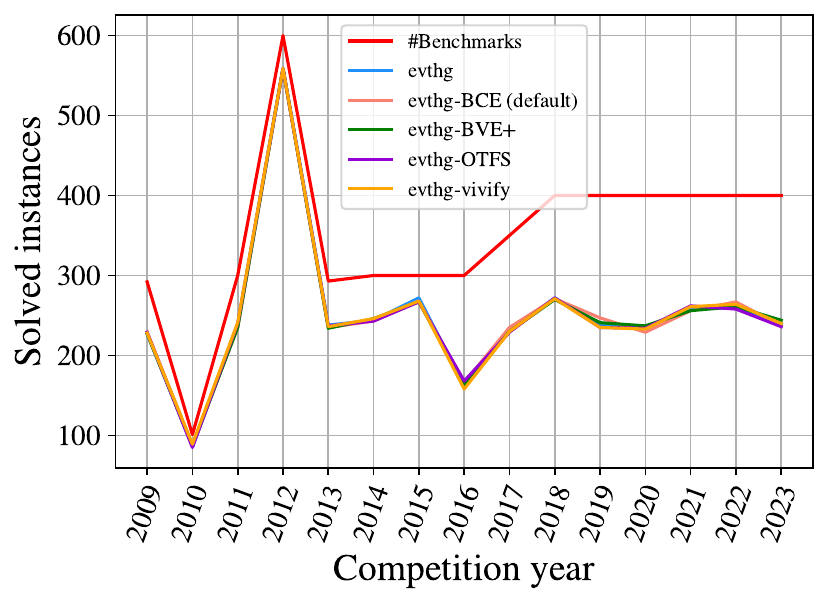}
    \caption{Comparison with everything}
    \label{fig:ev-all} 
  \end{subfigure}%
  \caption{Competition Results of settings compared to base/default/everything version of CaDiCaL}
  \label{fig:all}
  \end{figure}

\begin{figure}[ht!]
    \centering
    \begin{subfigure}{.5\textwidth}
      \centering
      \includegraphics[height=.7\linewidth]{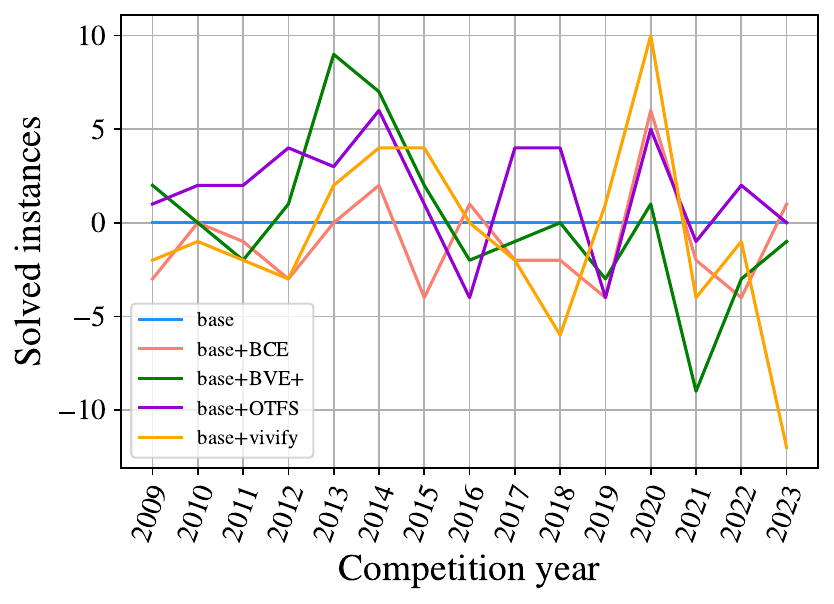}
      \caption{Difference to base}
      \label{fig:base-all-diff}
    \end{subfigure}
    \begin{subfigure}{.5\textwidth}
      \centering
      \includegraphics[height=.7\linewidth]{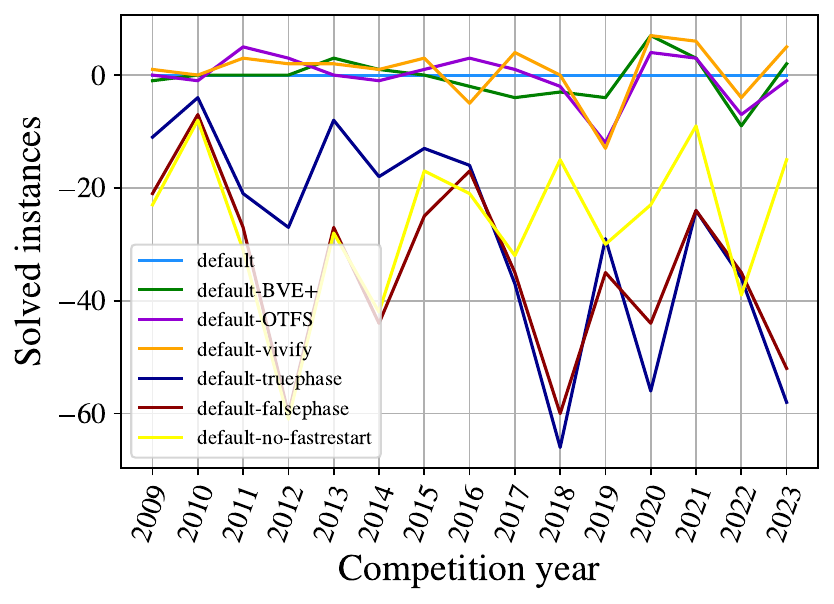}
      \caption{Difference to default}
      \label{fig:def-all-diff}
    \end{subfigure}%
    \begin{subfigure}{.5\textwidth}
      \centering
      \includegraphics[height=.7\linewidth]{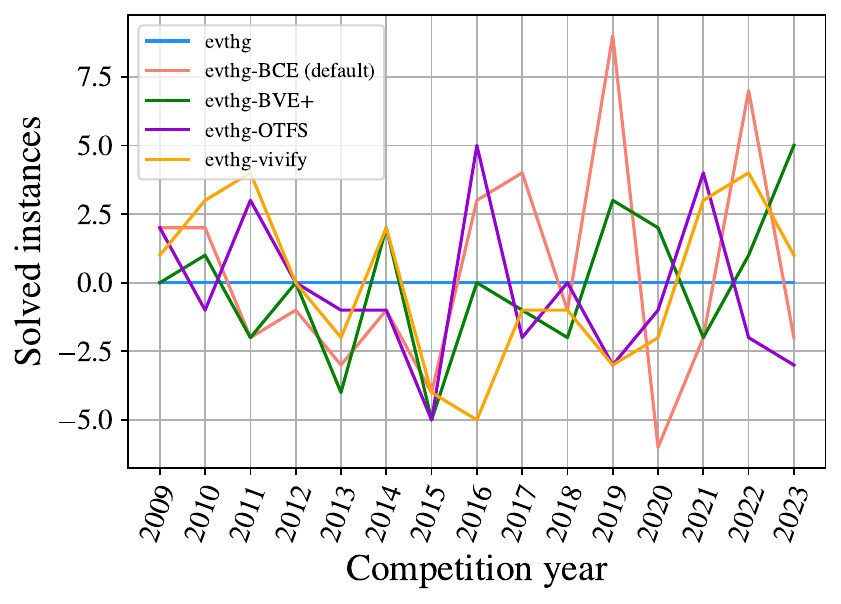}
      \caption{Difference to everything}
      \label{fig:ev-all-diff}
    \end{subfigure}
    \caption{Difference in solved instances of settings compared to default/base version of CaDiCaL}
    \label{fig:all-diff}
    \end{figure}

We provide a graphical interpretation of the three experiments in Figure~\ref{fig:all} and Figure~\ref{fig:all-diff}.
Figure~\ref{fig:all} depicts the number of solved instances, and we see that there are some differences in between the versions. However they are generally not so huge, that we can detect a clear trend.

  Hence we report the changes on the solved instances in Figure~\ref{fig:all-diff}. 
  We calculate the difference of the versions to either the base version (Figure~\ref{fig:base-all-diff}), the default version (Figure~\ref{fig:def-all-diff}), and everything version (Figure~\ref{fig:ev-all-diff}). 

  In Figure~\ref{fig:base-all-diff} we do not see a clear tendency of the spikes. For example in the year 2014 enabling individual features solves more instances than the base version of CaDiCaL. However in 2019 this is not the case. In general, BCE solves fewer instances in nearly every case, despite of 2014, 2016, and 2020. Hence, we believe that Hypothesis 2 is wrong. 

Figure~\ref{fig:def-all-diff} shows that we have a range of roughly +/-10 instances that can be solved. 
Interestingly all versions that are different from the default setting of CaDiCaL behave almost the same, i.e, they are either all above or all below the blue line. 
For example in the year 2019, all versions solve less instances than the default setting. This was unexpected for us because activating any feature in the base version does not yield a performance increase in this year. 
Hence we conclude that Hypothesis 1 is wrong. 

In Figure~\ref{fig:ev-all-diff} we do not see such a clear trend as in Figure~\ref{fig:def-all-diff}. The different versions are much more spread around the everything-version (blue line). In the years 2013 and 2015 disabling a single feature lead to a drop in the solved instances. In the remaining years some versions with a single feature being disabled were able to solve more instances, others solved less instances. 

There is also no correlation between the techniques: neither in the base version nor in the default or everything experiment, they do not follow each other. This contradicts our third hypothesis.


In the following sections we provide more detailed results on the individual features.
Additionally we have generated cactus plots for each individual year of each experiment. The cactus plots can be seen in Appendices~\ref{app:cactus-base}--\ref{app:cactus-everything}.
 
\subsection{BCE}

\begin{figure}[ht!] 
    \centering
    \begin{subfigure}{.5\textwidth}
      \centering
      \includegraphics[height=.7\linewidth]{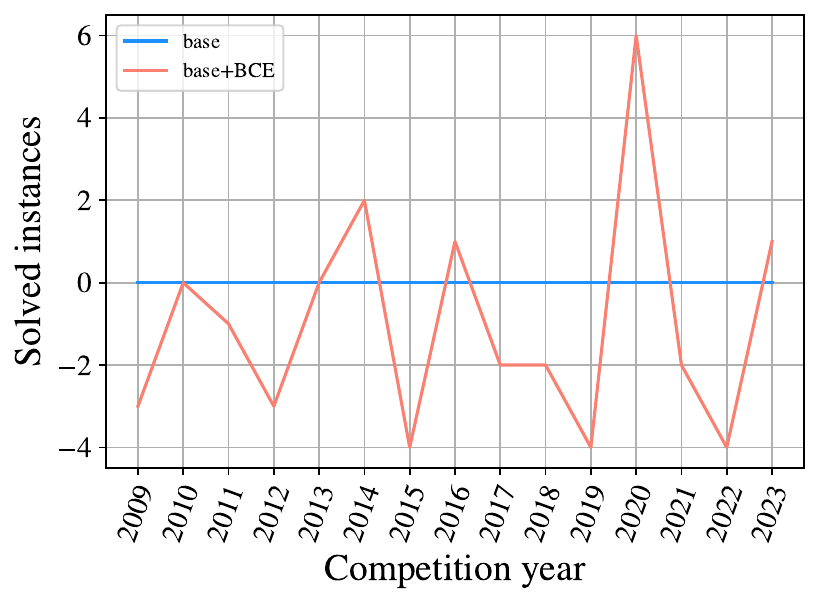}
      \caption{Difference to base}
      \label{fig:base-bce-diff}
    \end{subfigure}%
    \begin{subfigure}{.5\textwidth}
      \centering
      \includegraphics[height=.7\linewidth]{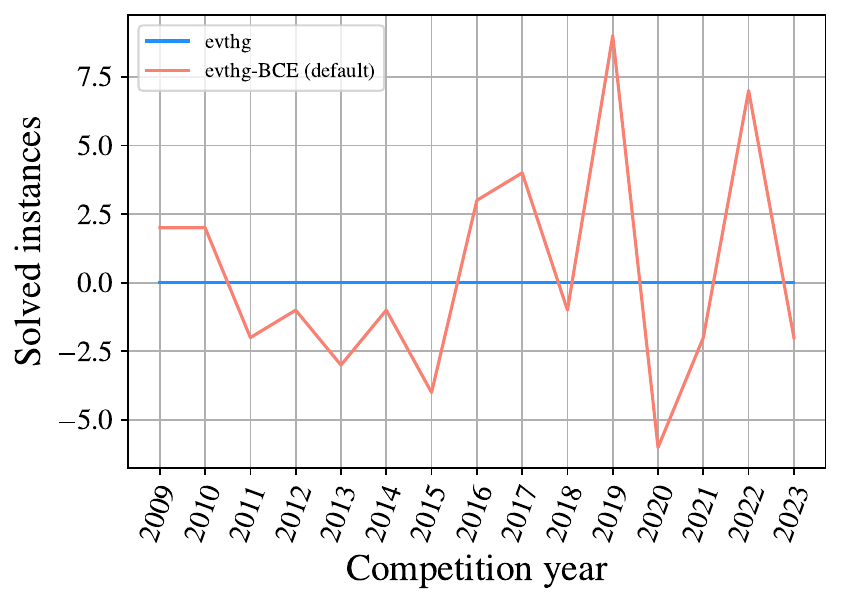}
      \caption{Difference to everything}
      \label{fig:ev-bce-diff}
    \end{subfigure}
    \caption{Detailed results for BCE} 
    \label{fig:bce}
    \end{figure}

Blocked clause elimination is turned off by default in CaDiCaL. Hence it is not included in the experiment ``default''.
In general, enabling only BCE performs almost always poorer than the base version of CaDiCaL, as can be seen in Figure~\ref{fig:base-bce-diff}. In only three years adding BCE lead to an increase in the solved instances.

Removing BCE from the everything-version can be seen in Figure~\ref{fig:ev-bce-diff}.
In the year 2019 disabling BCE lead to 9 more solved instances, whereas 6 less instances where solved in 2020.

  \subsection{OTFS}
\begin{figure}[ht!]
      \centering
        \begin{subfigure}{.5\textwidth}
          \centering
          \includegraphics[height=.7\linewidth]{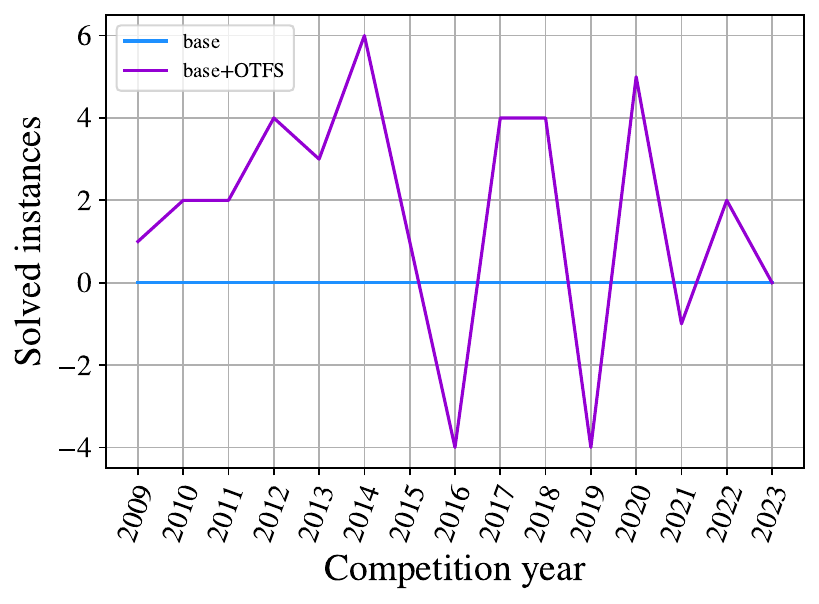}
          \caption{Difference to base}
          \label{fig:base-otfs-diff}
        \end{subfigure}
        \begin{subfigure}{.5\textwidth}
          \centering
        \includegraphics[height=.7\linewidth]{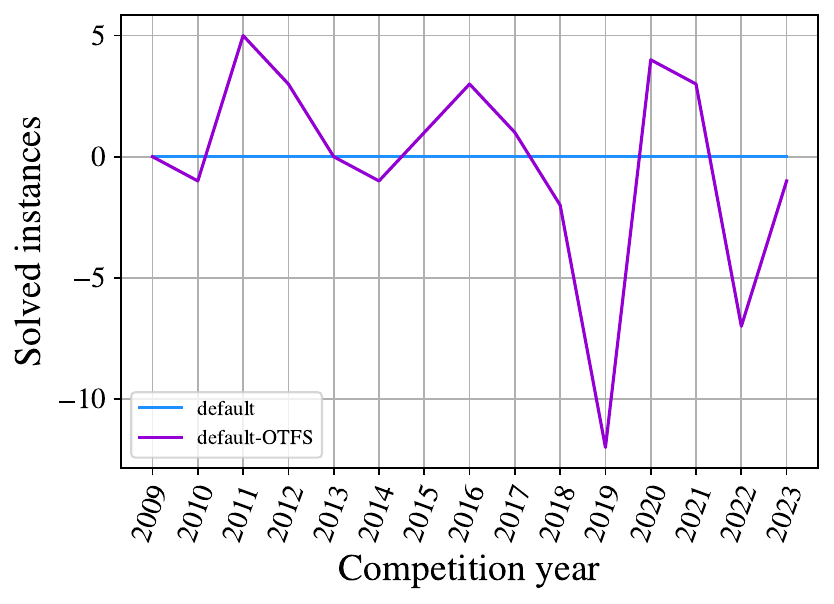}
        \caption{Difference to default}
        \label{fig:def-otfs-diff}
      \end{subfigure}%
      \begin{subfigure}{.5\textwidth}
        \centering
        \includegraphics[height=.7\linewidth]{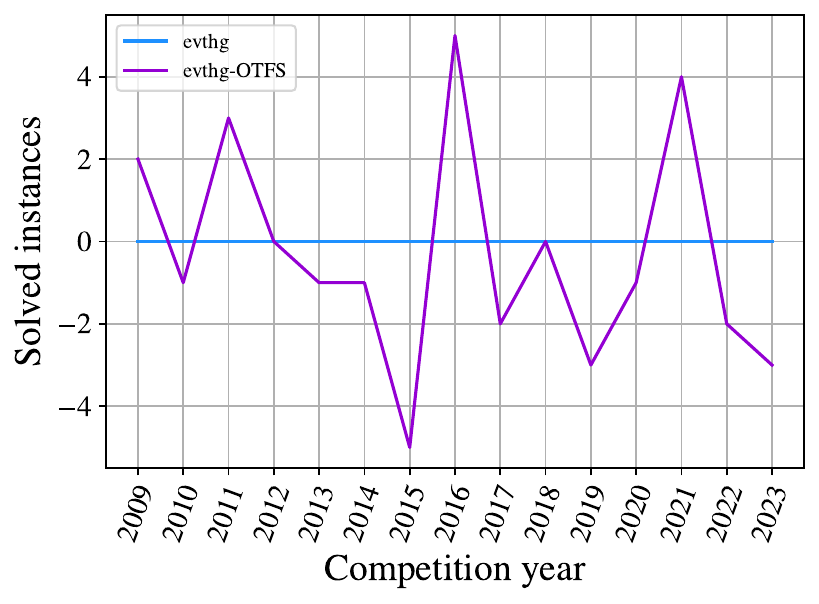}
        \caption{Difference to everything}
        \label{fig:ev-otfs-diff}
      \end{subfigure}
      \caption{Detailed results for  OTFS}
      \label{fig:otfs}
      \end{figure}

  Enabling OTFS in the base experiment was beneficial in 11 years, as can be seen in Figure~\ref{fig:base-otfs-diff}.

  In Figure~\ref{fig:def-otfs-diff} we disable OTFS. Hence CaDiCaL only applies \bve and vivification. 
In the year 2011 disabling OFTS gave us +5 instances, whereas we lose 11 instances in the year 2019. 
In the experiment ``everything'' (Figure~\ref{fig:ev-otfs-diff}) disabling OTFS decreased the number of solved instances in 9 years.

\subsection{\bve{}}   

\begin{figure}[ht!]
        \centering
          \begin{subfigure}{.5\textwidth}
            \centering
            \includegraphics[height=.7\linewidth]{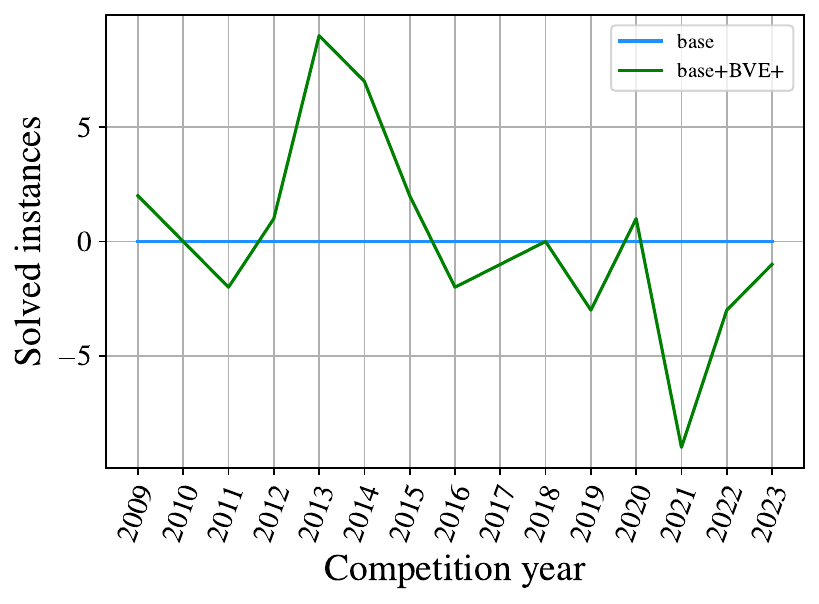}
            \caption{Difference to base}
            \label{fig:base-bve-diff}
          \end{subfigure}
          \begin{subfigure}{.5\textwidth}
            \centering
          \includegraphics[height=.7\linewidth]{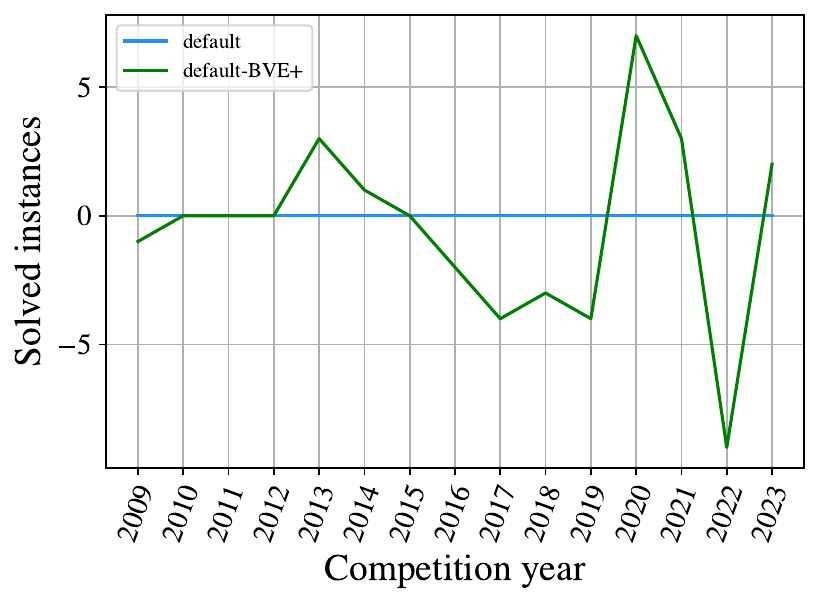}
          \caption{Difference to default}
          \label{fig:def-bve-diff}
        \end{subfigure}%
        \begin{subfigure}{.5\textwidth}
          \centering
          \includegraphics[height=.7\linewidth]{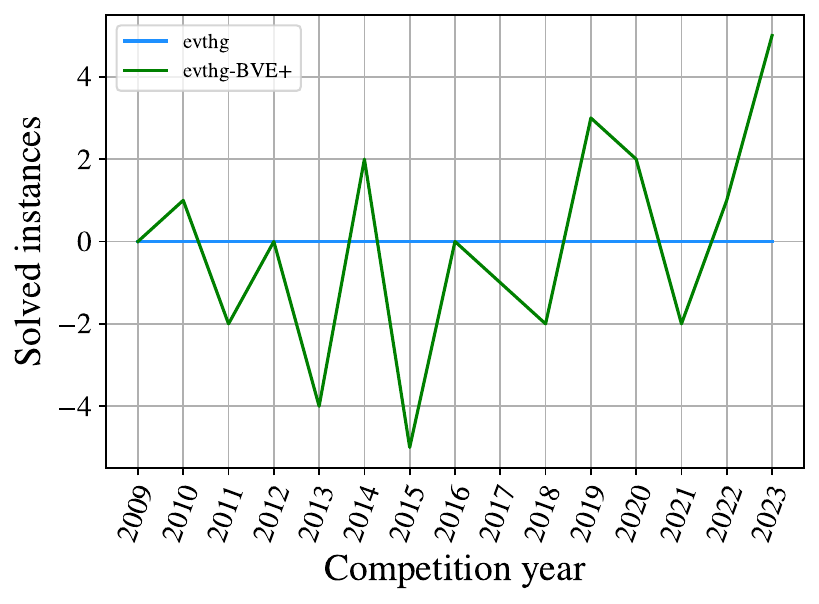}
          \caption{Difference to everything}
          \label{fig:ev-bve-diff}
        \end{subfigure}
        \caption{Detailed results for  \bve{}}
        \label{fig:bve}
        \end{figure}

Enabling \bve{} is an interesting case. In the years up to 2015 we see a benefit of enabling the technique. However 2016 acted as a turning point, and in the last couple of years enabling BVE+ actually decreases the number of solved instances (see Figure~\ref{fig:base-bve-diff}). 

Disabling \bve{} in the default and in the everything experiment lead to fewer solved instances in more than half of the years.

\subsection{Vivification} 

\begin{figure}[ht!]
          \centering
          \begin{subfigure}{.5\textwidth}
            \centering
            \includegraphics[height=.7\linewidth]{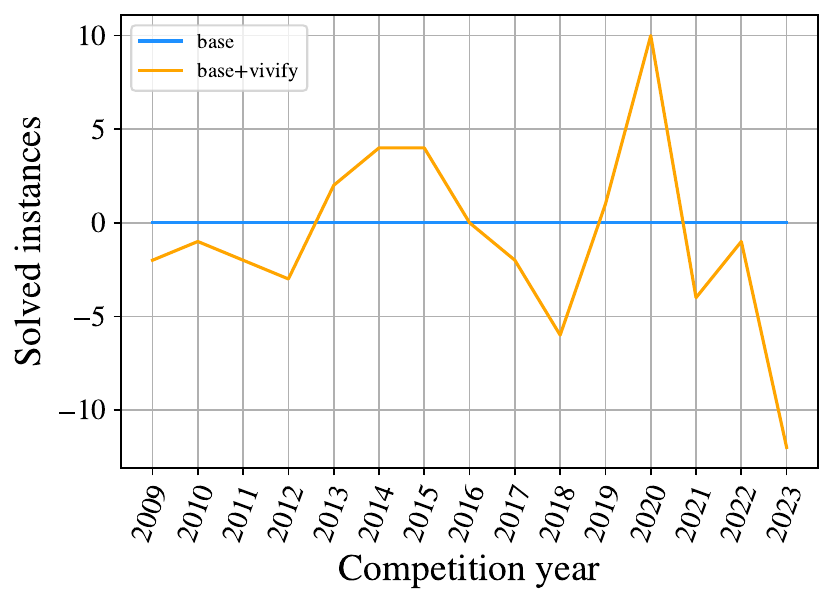}
            \caption{Difference to base}
            \label{fig:base-vivify-diff}
          \end{subfigure}
          \begin{subfigure}{.5\textwidth}
          \centering
          \includegraphics[height=.7\linewidth]{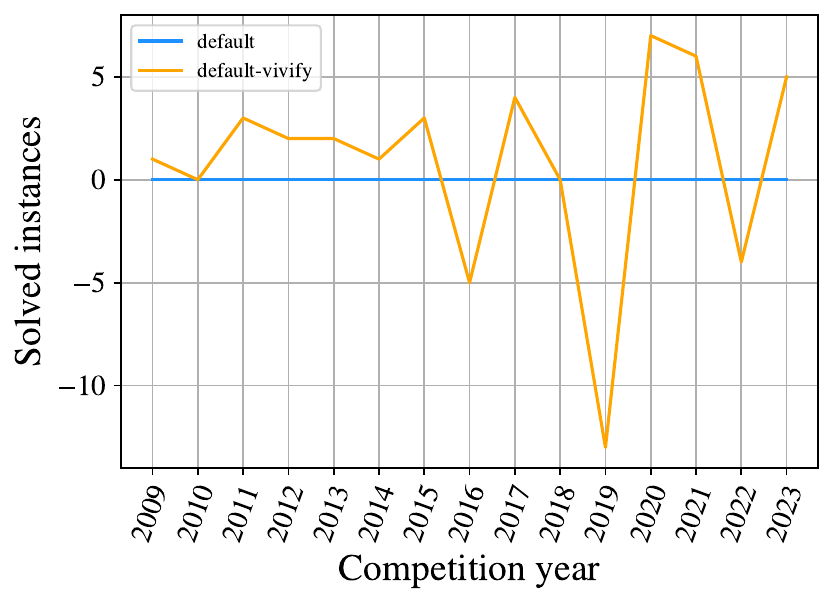}
          \caption{Difference to default}
          \label{fig:def-vivify-diff}
        \end{subfigure}%
          \begin{subfigure}{.5\textwidth}
            \centering
            \includegraphics[height=.7\linewidth]{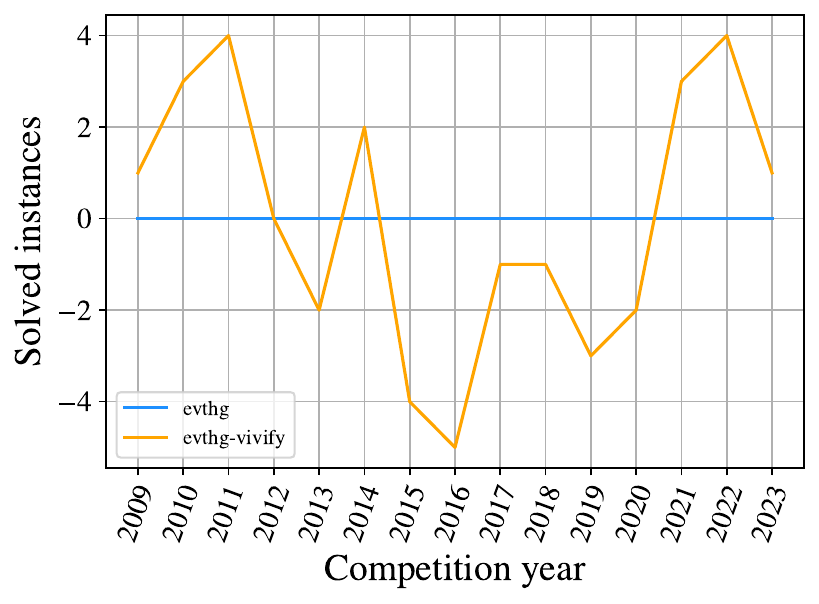}
            \caption{Difference to everything}
            \label{fig:ev-vivify-diff}
          \end{subfigure}
          \caption{Detailed results for  vivify}
          \label{fig:vivify}
          \end{figure}

When all features are disabled, vivification neither shows a clear benefit nor disadvantage as can be seen in Figure~\ref{fig:base-vivify-diff}.
However, when considering the default configuration, it seems not to be very useful, except for the SAT Competition 2019 where
vivification was very successful, see Figure~\ref{fig:def-vivify-diff}.

\subsection{Interdependences of the Techniques}

\begin{figure}[ht!]
          \centering
          \begin{subfigure}{.5\textwidth}
            \centering
            \includegraphics[height=.7\linewidth]{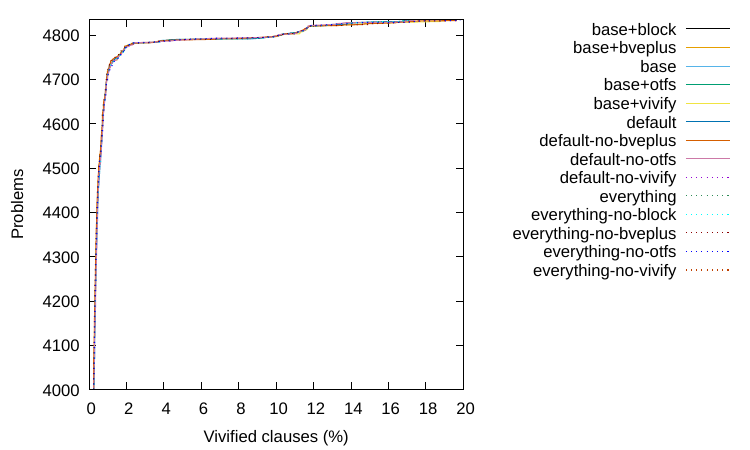}
            \caption{CDF of the efficiency of vivification (number of changed clauses: strengthened or eliminated)}
            \label{fig:stats-vivify-diff}
          \end{subfigure}
          
          \begin{subfigure}{.5\textwidth}
          \centering
            \includegraphics[height=.7\linewidth]{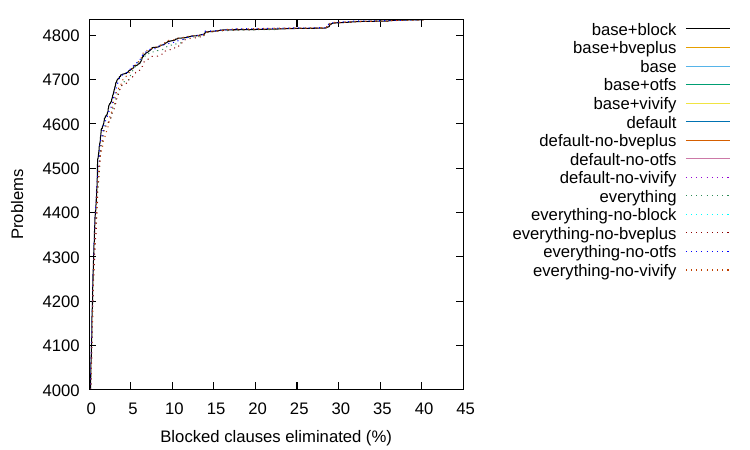}
          \caption{CDF of the efficiency of BCE (eliminated irredundant clause)}
          \label{fig:stats-blocked-diff}
        \end{subfigure}
        
          \begin{subfigure}{.5\textwidth}
            \centering
            \includegraphics[height=.7\linewidth]{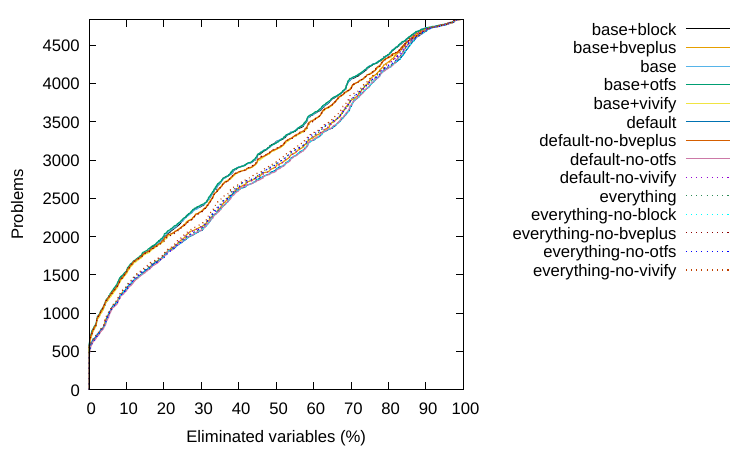}
            \caption{CDF of the efficiency of \bve (eliminated variables in \% of the overall overall variables)}
            \label{fig:stas-elim-diff}
          \end{subfigure}
          \caption{Efficiency of various techniques over all problems, including unsolved ones}
          \label{fig:stats-diff}
          \end{figure}

We have slightly scratched the surface of the information provided by
our log files. We looked at how often techniques are applied
successfully (according to the output of CaDiCaL). For our techniques
this amount to check how many variables are eliminated (where we
expect \emph{more} variables to be eliminated with \bve), the number
of clauses that have been changed by vivification, and finally the
number of clauses removed by BCE. To compare the numbers, we compare
the different percentages over the \emph{entire} set of problems,
including unsolved ones.  To visualize the results, we show CDFs for
all the problems (Fig.~\ref{fig:stats-diff}): the more to the right
and the more to the bottom, the more effective a technique is.  Remark
that a technique can be successful without any change on the future
run of the solver or even making the performance worse. We are here
only interested in how often a technique was successful.

For vivification and BCE, we observe very little difference between
the configuration (when the option is activated): the curves are nearly
identical.

For \bve, we do observe a difference (Fig.\ref{fig:stas-elim-diff}).
There are three group of curves: (i) the base curves (without
vivification), (ii) the techniques with vivification but without \bve,
and (iii) the others. This matches our intuition that vivification
strengthen (shortens) clauses, making more eliminations possible and
that \bve eliminates more variables than not activating it. However,
we expected to see a difference between everything (with BCE) and
default (without BCE): BCE is a technique to remove even more
clauses, which in theory could enable more eliminations, but this
does not seem to be the case.

Overall, for the techniques we have considered, it seems that the
techniques are independent enough of each other we cannot observe a
difference.

\section{Interpretation and Conclusion}\label{sec:conclusion}

Regarding our hypotheses of Section~\ref{sec:expectations}, we do not see a constant drop of any feature in the default-experiment. In some years disabling certain features increases the number of solved instances, in other years we loose instances. So we cannot confirm Hypothesis 1. 

In the base case we also do not see our estimated spikes. In some cases, this could be due to implementation details: the recent SAT Competition winner Kissat has a more advanced scheduling approach for vivification. But even in Kissat vivification of irredundant clauses has been activated (2020), deactivated, and reactivated (2023). Similar criticism can be given to our OTFS implementation: it does not change the bumped variables when changing the conflict. We cannot confirm Hypothesis 2 either.

Furthermore, no two features show exactly the same behaviour for all years. Hence we do not see the effect that one technique simulates another feature. 
Hypothesis 3 remains unconfirmed. 

\paragraph{Conclusion.} Since all our expectations remain unconfirmed, we come to the conclusion that it is not a good idea to single out features and test them with either being enabled or disabled. We think the features influence each other too much. To give a clear rejection of our hypothesis for the default setting, we would need to do some kind of causal analysis or Monte-Carlo grid method to determine the influence the features have on each other. 
However, this exceeded the computation time we had for this paper. Another way to go deeper in the analysis is to
look at the statistics. Besides overall solving time, it would be interesting to check how successful
various techniques are.

\begin{acknowledgments}
   This work is supported by the Austrian Science Fund (FWF), NFN
  S11408-N23 (RiSE) and the LIT AI Lab funded by the State of Upper
  Austria.
\end{acknowledgments}
\bibliography{paper}
\newpage
\appendix
\section{Cactus Plots - Experiment ``base''}
\label{app:cactus-base}

\begin{figure}[ht!]
  \centering
  \begin{subfigure}{.5\textwidth}
    \centering
    \includegraphics[height=.6\linewidth]{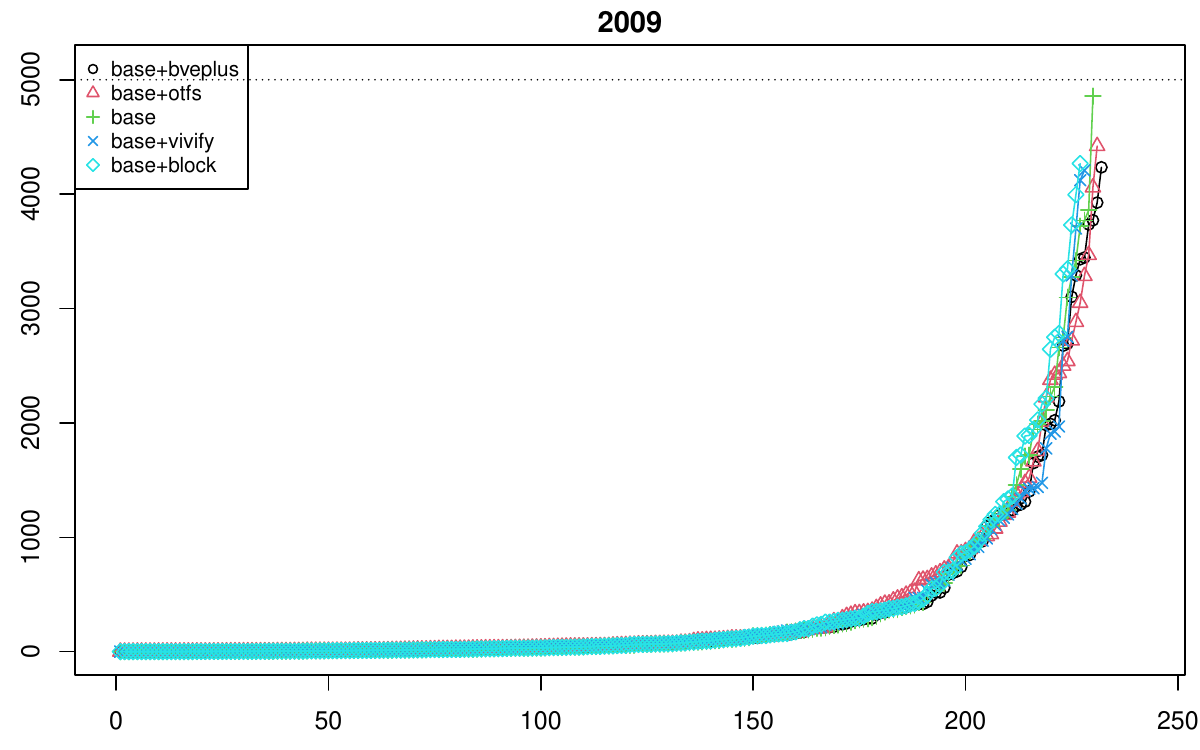}
  \end{subfigure}%
  \begin{subfigure}{.5\textwidth}
    \centering
    \includegraphics[height=.6\linewidth]{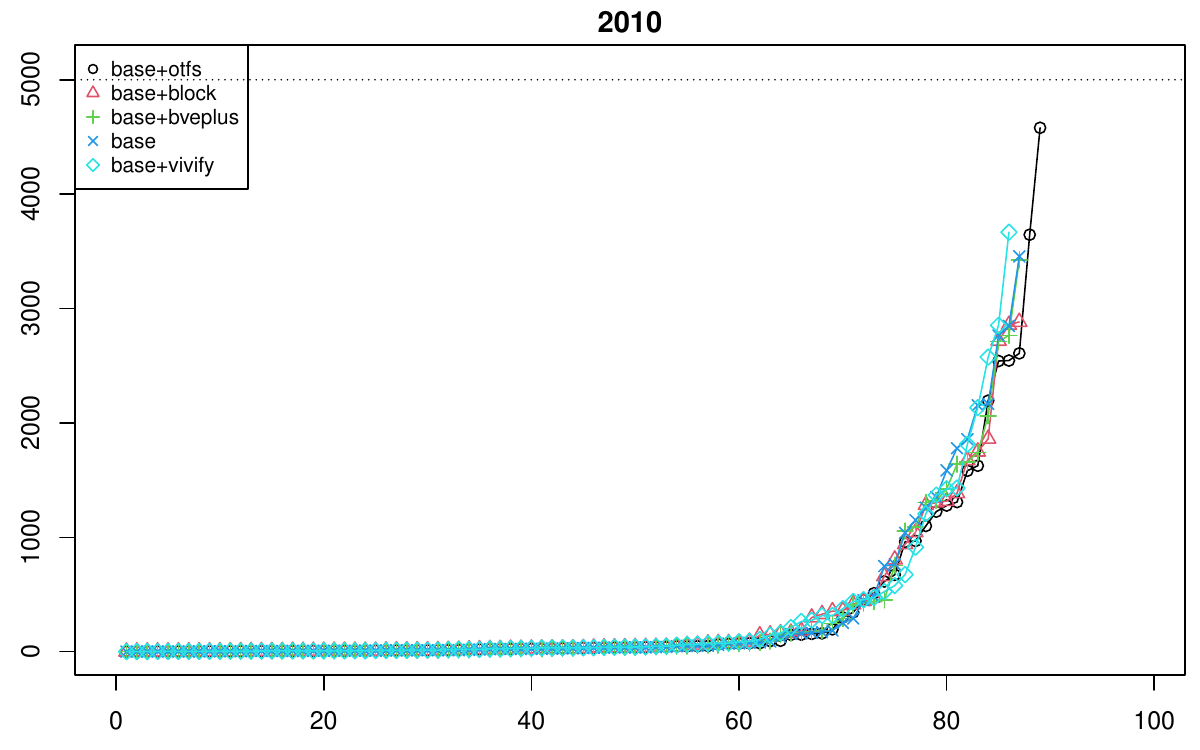}
  \end{subfigure}
  \begin{subfigure}{.5\textwidth}
    \centering
    \includegraphics[height=.6\linewidth]{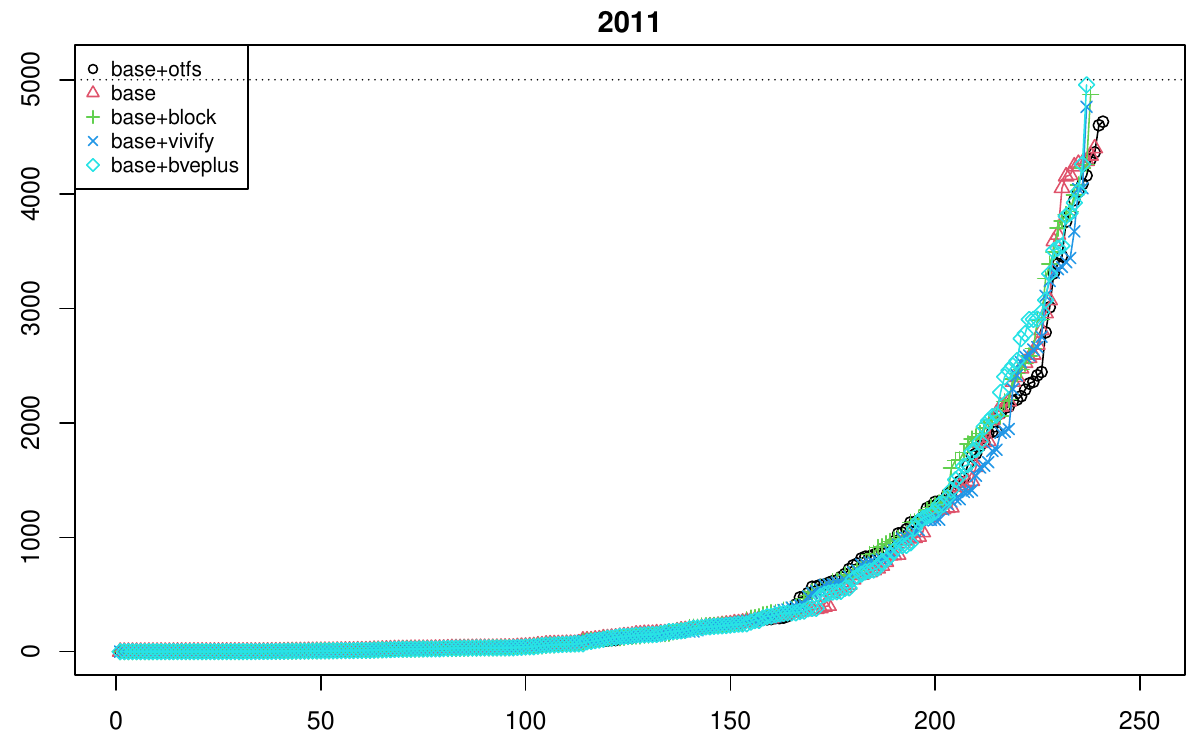}
    \end{subfigure}%
  \begin{subfigure}{.5\textwidth}
    \centering
    \includegraphics[height=.6\linewidth]{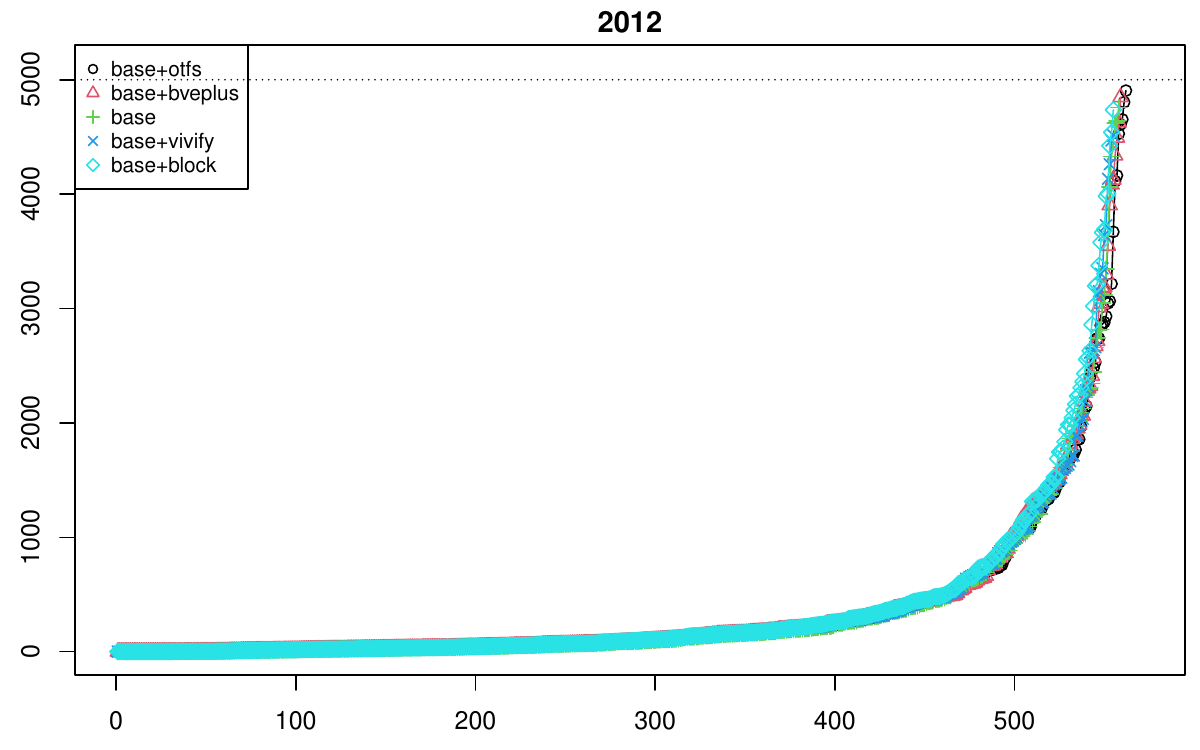}
    \end{subfigure}
  \begin{subfigure}{.5\textwidth}
    \centering
    \includegraphics[height=.6\linewidth]{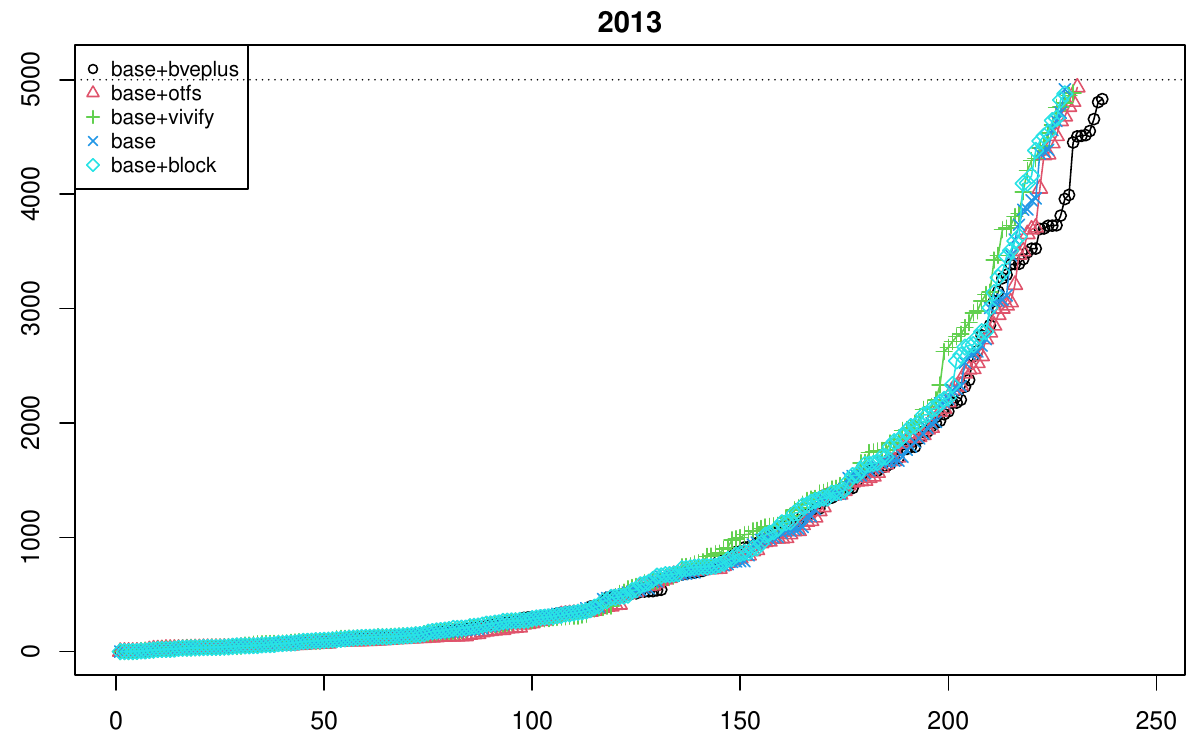}
  \end{subfigure}%
  \begin{subfigure}{.5\textwidth}
    \centering
    \includegraphics[height=.6\linewidth]{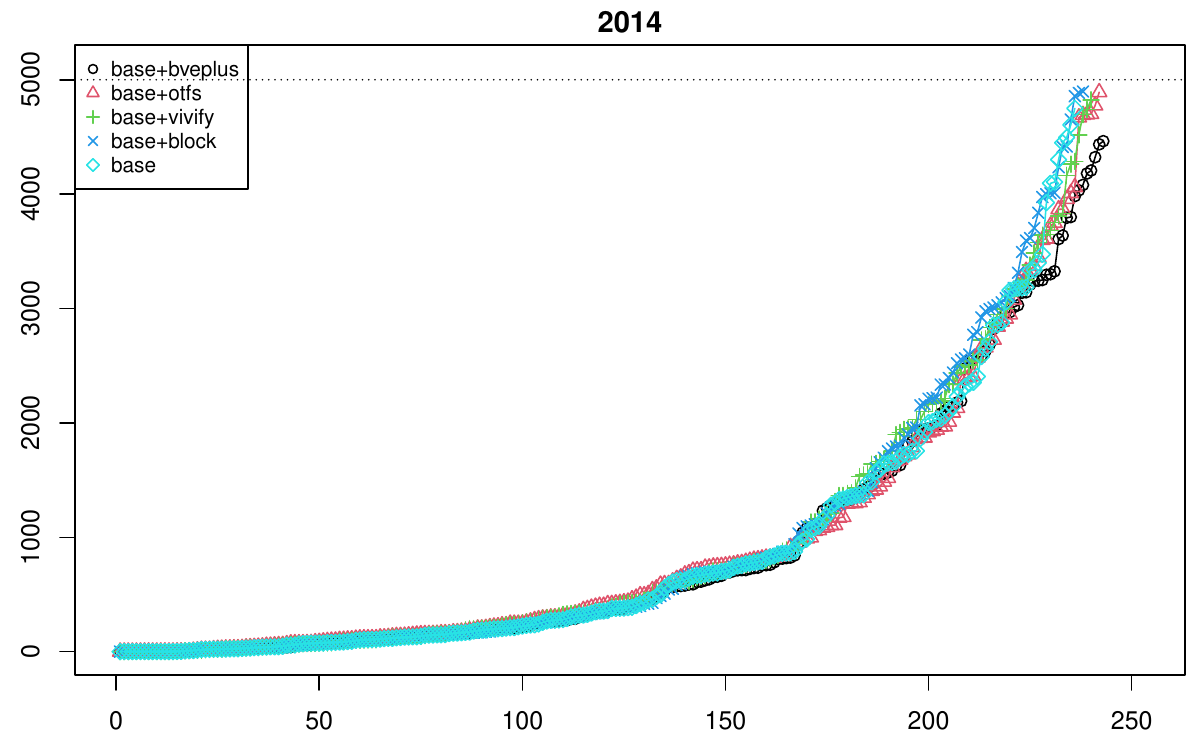}
   \end{subfigure}
  \begin{subfigure}{.5\textwidth}
    \centering
    \includegraphics[height=.6\linewidth]{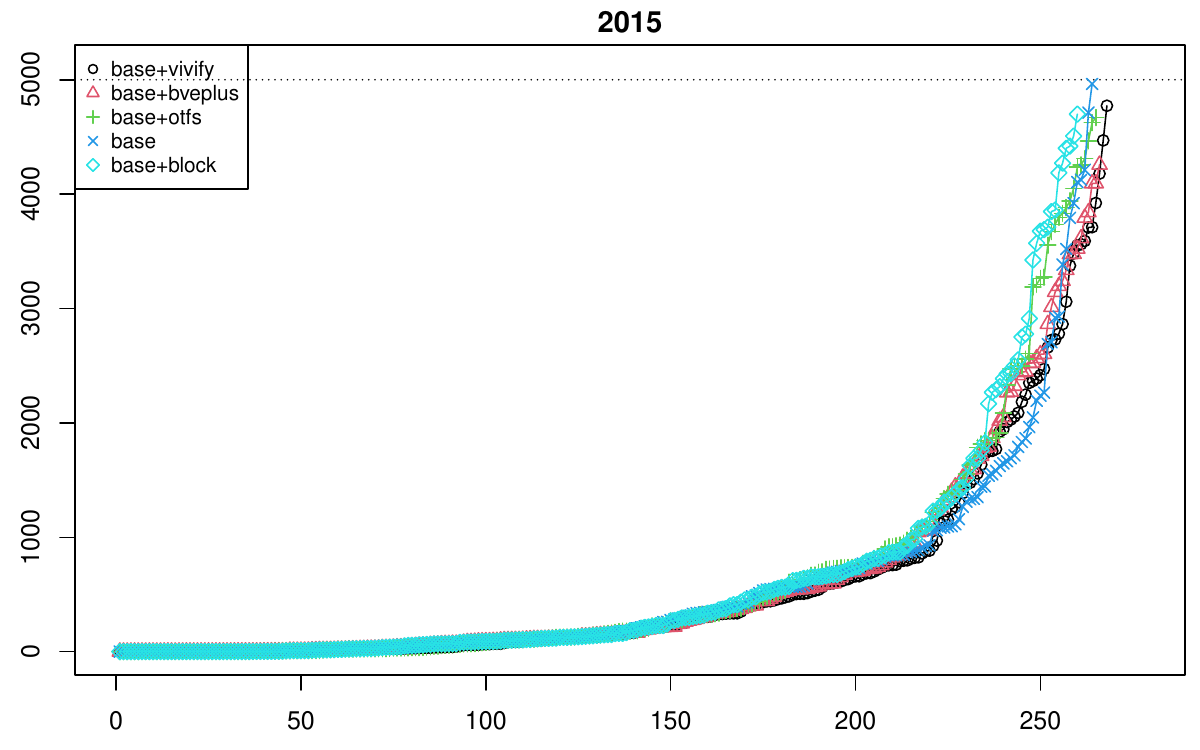}
   \end{subfigure}%
  \begin{subfigure}{.5\textwidth}
    \centering
    \includegraphics[height=.6\linewidth]{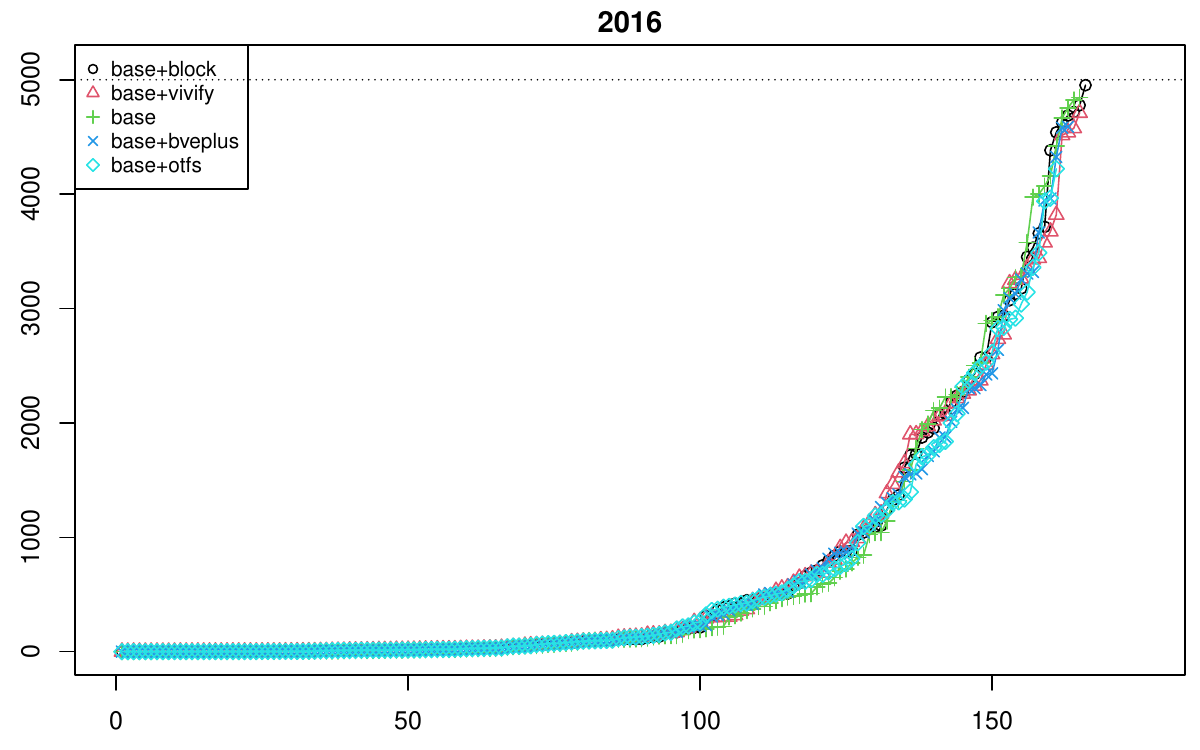}
   \end{subfigure}
   \caption{Cactus Plots of experiment ``base'' -- years 2009-2016}
  \end{figure}

  \begin{figure}[p]
    \centering
  \begin{subfigure}{.5\textwidth}
    \centering
    \includegraphics[height=.6\linewidth]{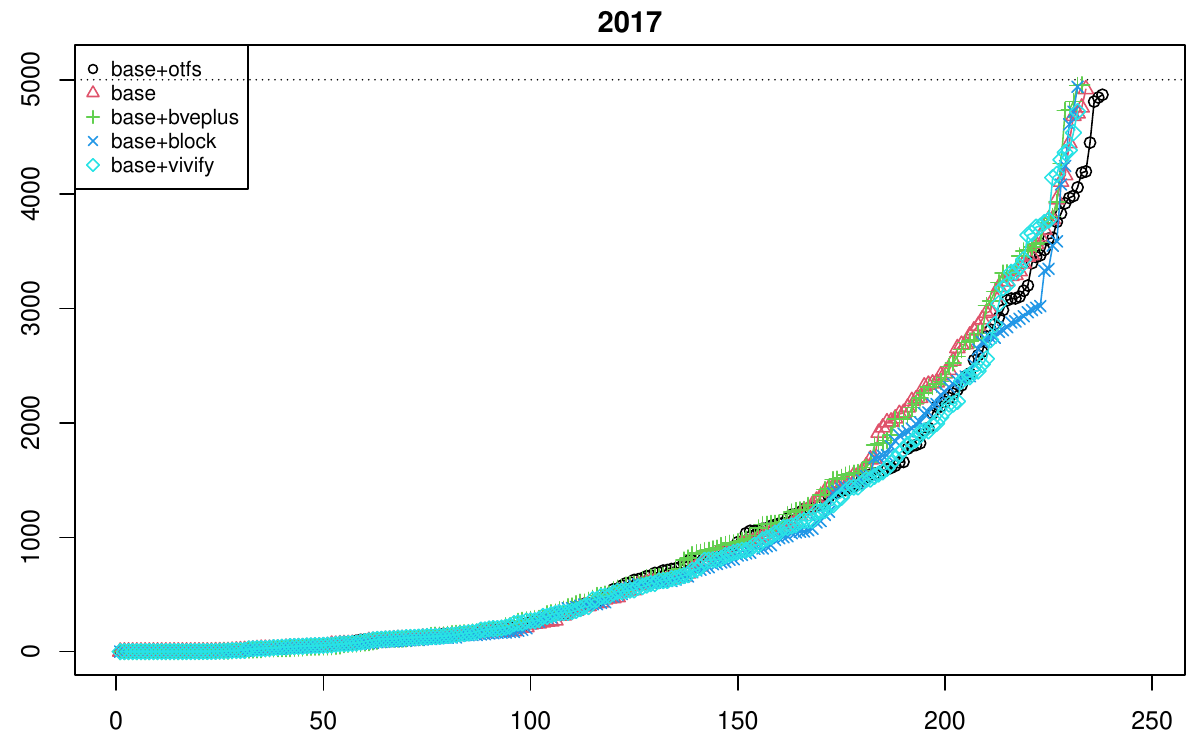}
   \end{subfigure}%
  \begin{subfigure}{.5\textwidth}
    \centering
    \includegraphics[height=.6\linewidth]{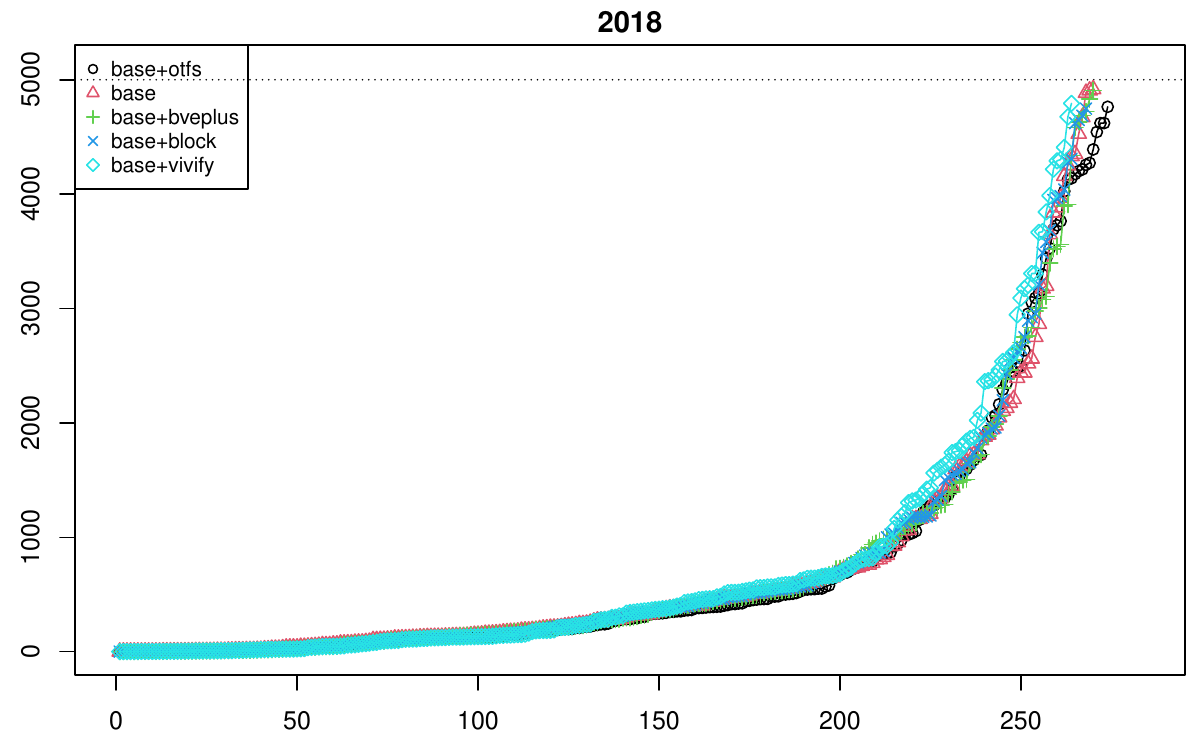}
   \end{subfigure}

  \begin{subfigure}{.5\textwidth}
    \centering
    \includegraphics[height=.6\linewidth]{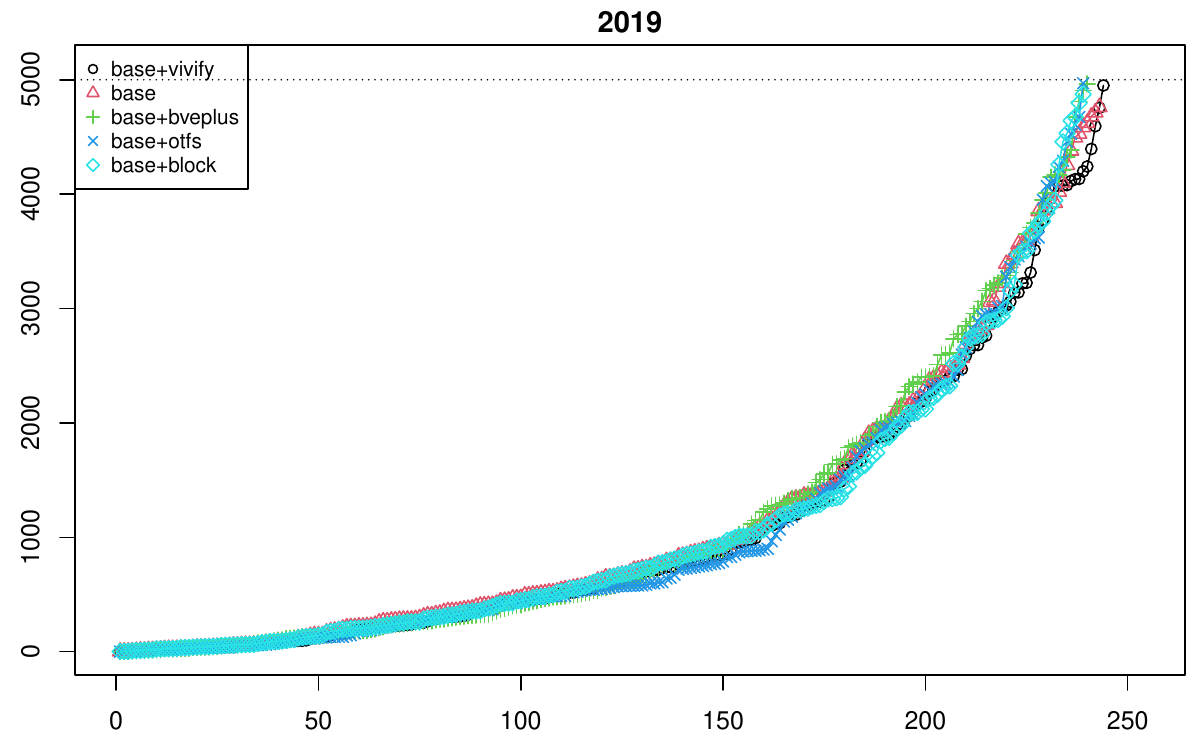}
  \end{subfigure}%
  \begin{subfigure}{.5\textwidth}
    \centering
    \includegraphics[height=.6\linewidth]{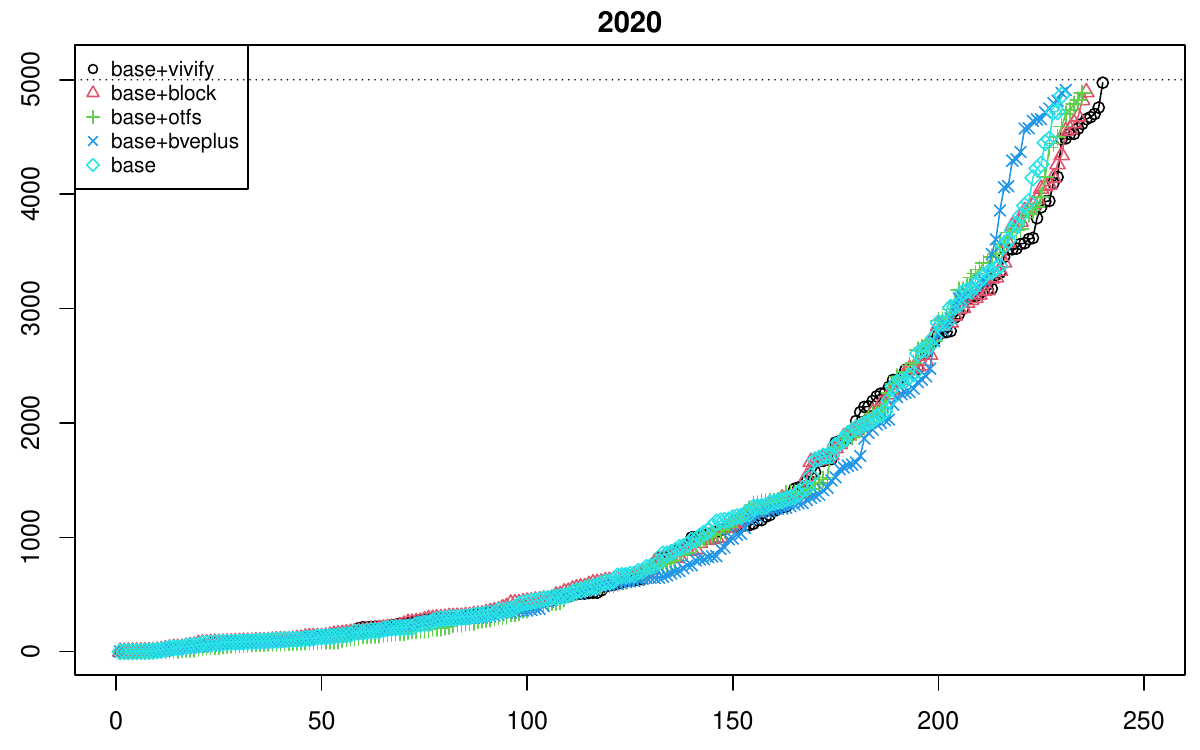}
  \end{subfigure}
  \begin{subfigure}{.5\textwidth}
    \centering
    \includegraphics[height=.6\linewidth]{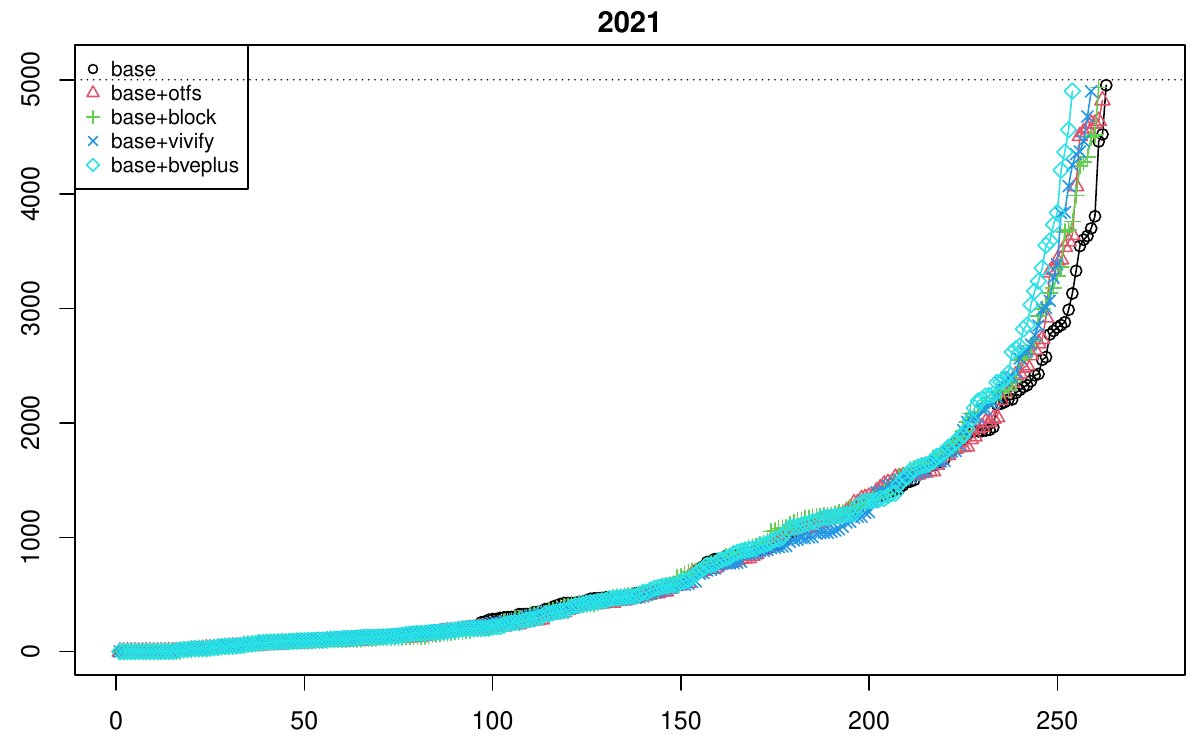}
  \end{subfigure}%
  \begin{subfigure}{.5\textwidth}
    \centering
    \includegraphics[height=.6\linewidth]{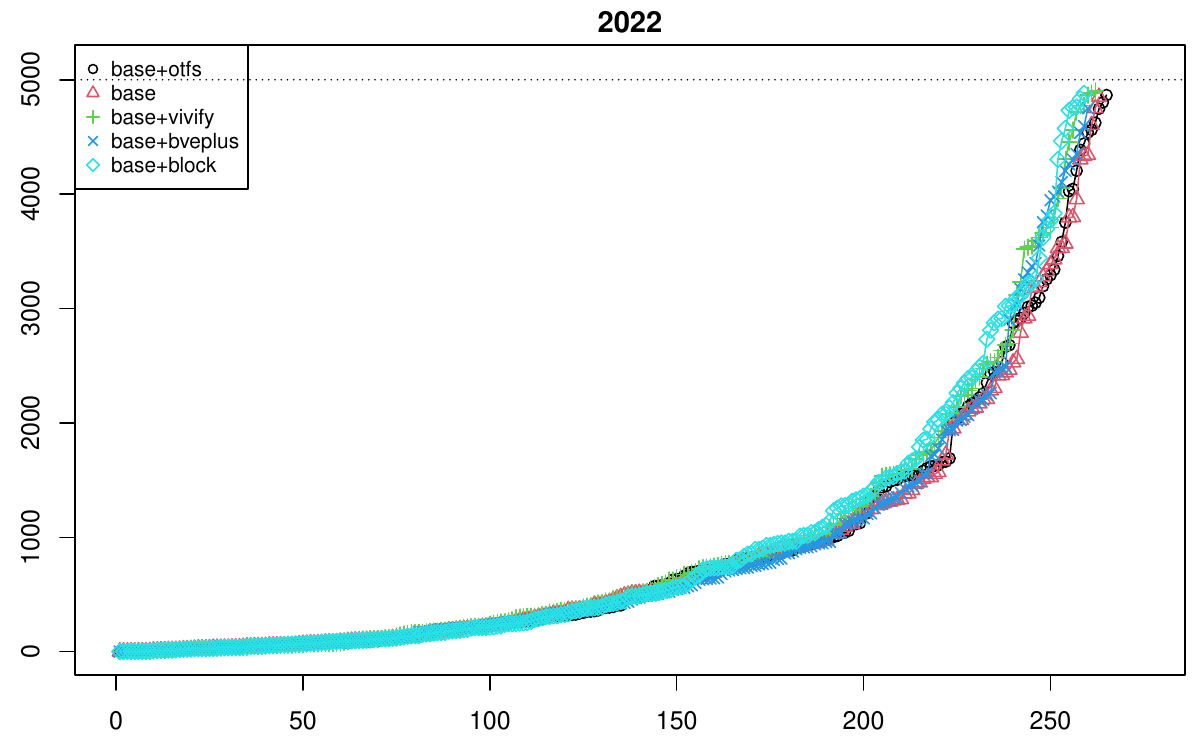}
   \end{subfigure}
  \caption{Cactus Plots of experiment ``base'' -- years 2017-2022}
  \end{figure}

\newpage
\section{Cactus Plots - Experiment ``default''}
\label{app:cactus-default}

  \begin{figure}[ht!]
    \centering
    \begin{subfigure}{.5\textwidth}
      \centering
      \includegraphics[height=.6\linewidth]{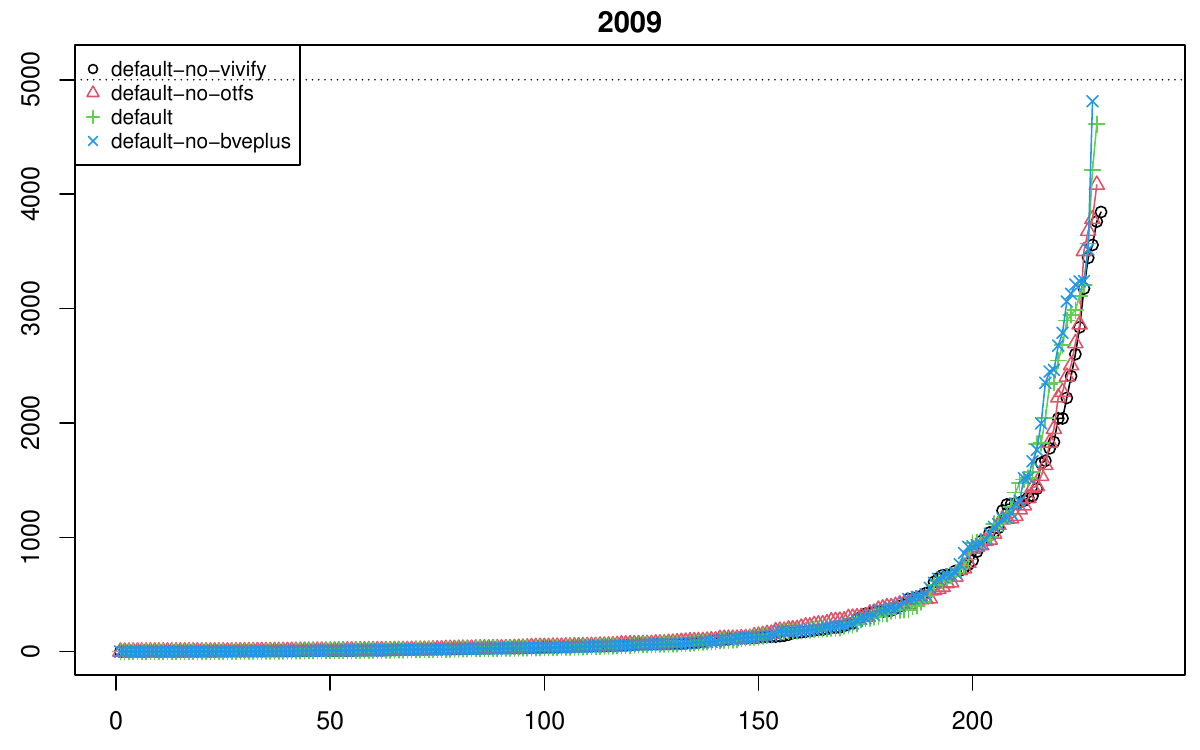}
    \end{subfigure}%
    \begin{subfigure}{.5\textwidth}
      \centering
      \includegraphics[height=.6\linewidth]{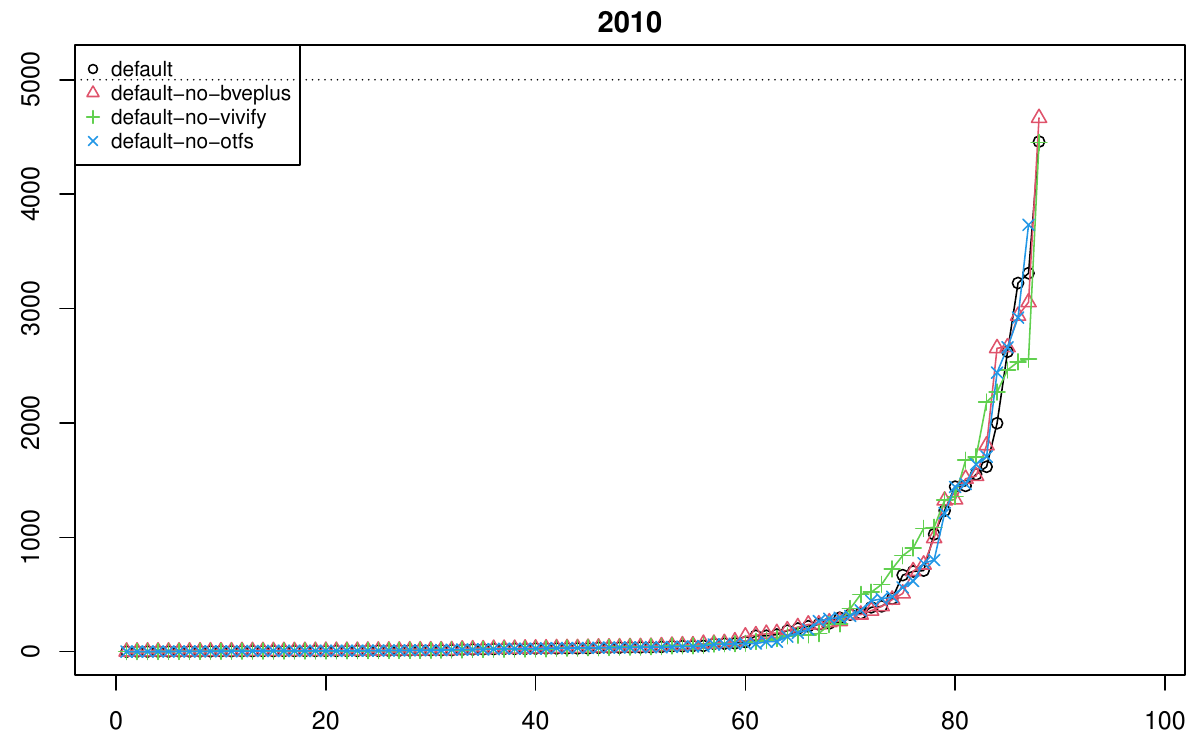}
    \end{subfigure}
    \begin{subfigure}{.5\textwidth}
      \centering
      \includegraphics[height=.6\linewidth]{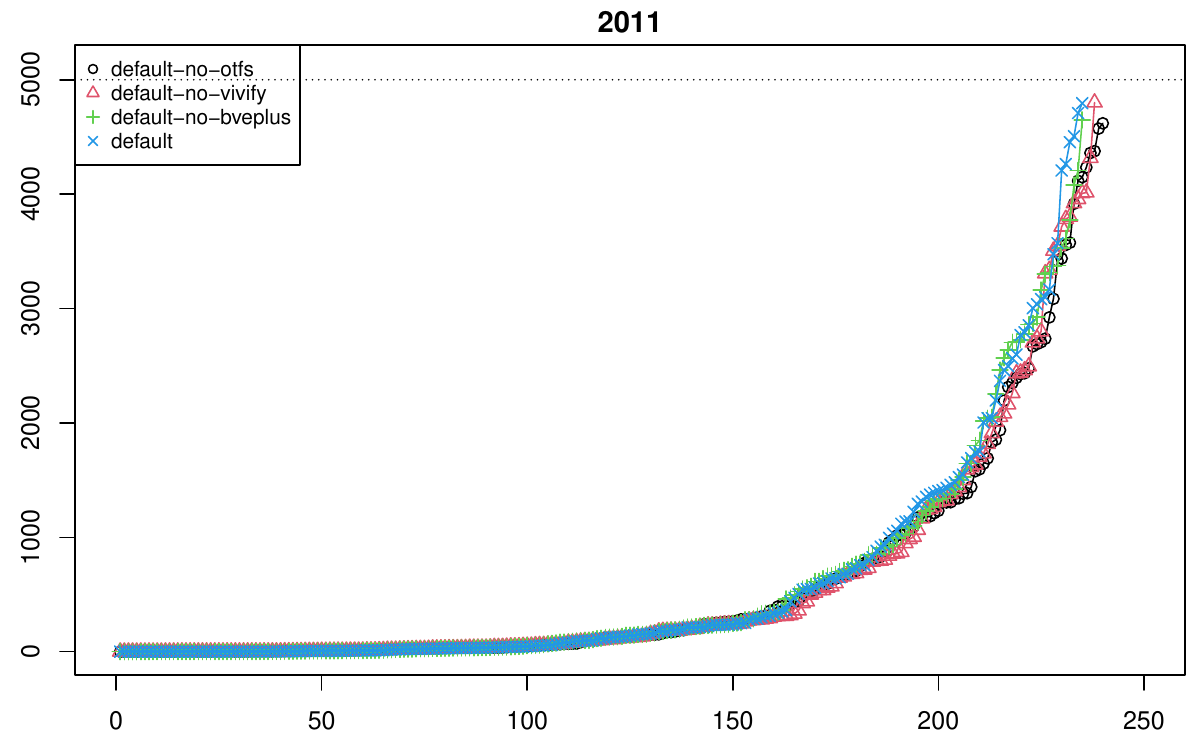}
      \end{subfigure}%
    \begin{subfigure}{.5\textwidth}
      \centering
      \includegraphics[height=.6\linewidth]{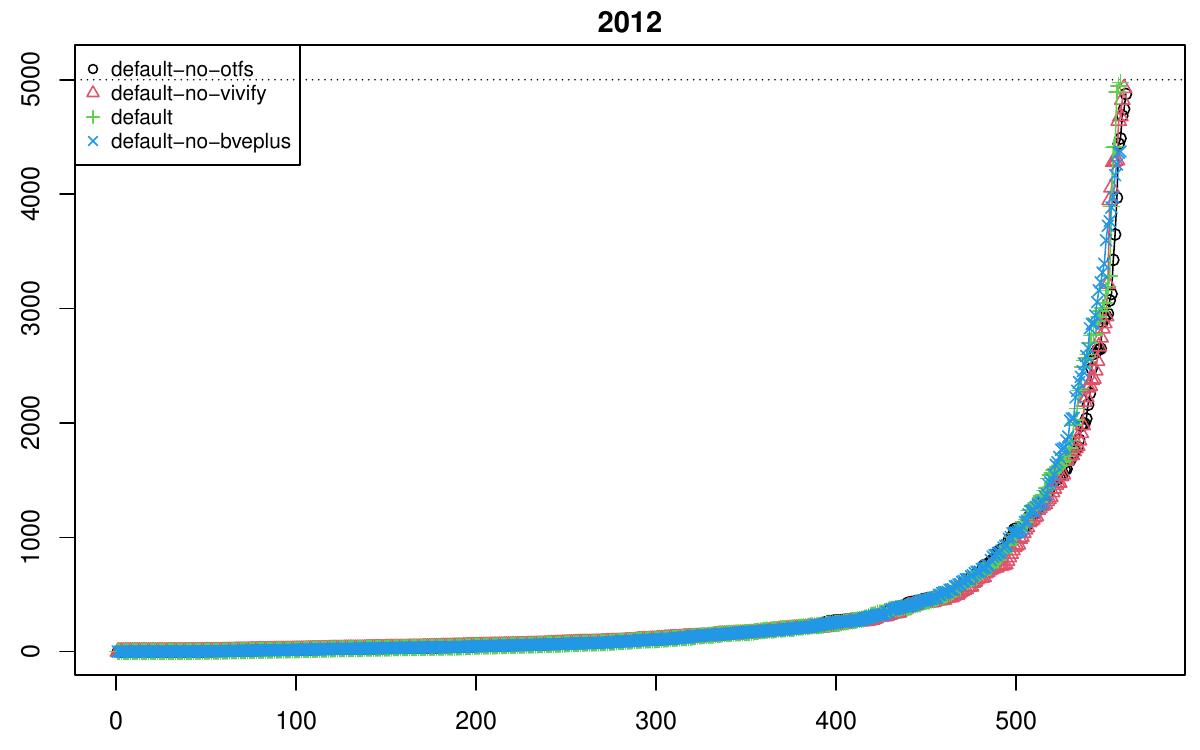}
      \end{subfigure}
     \begin{subfigure}{.5\textwidth}
      \centering
      \includegraphics[height=.6\linewidth]{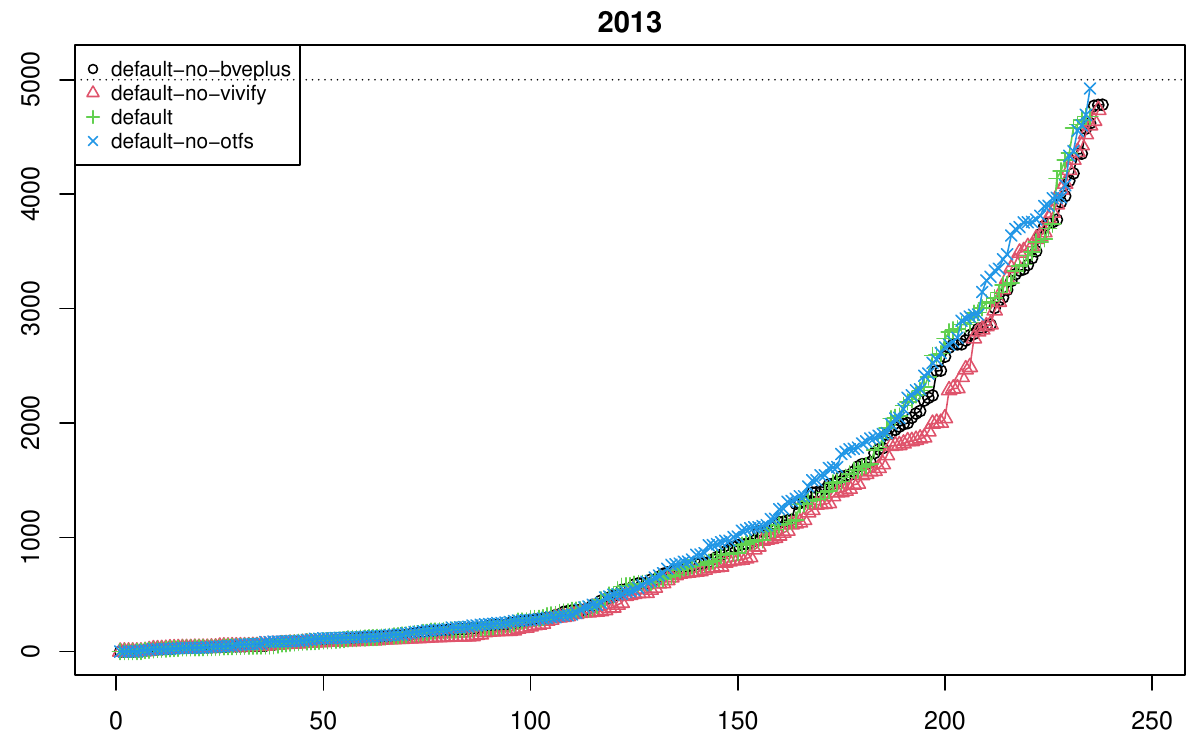}
    \end{subfigure}%
    \begin{subfigure}{.5\textwidth}
      \centering
      \includegraphics[height=.6\linewidth]{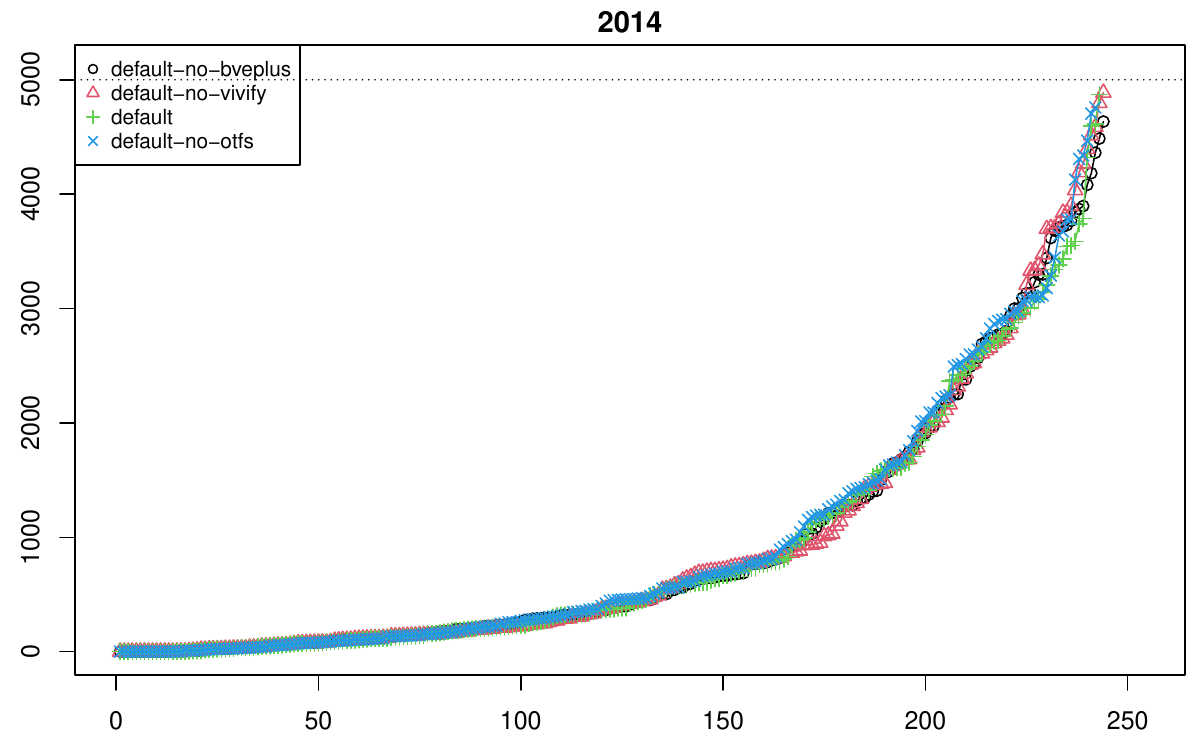}
     \end{subfigure}

    \begin{subfigure}{.5\textwidth}
      \centering
      \includegraphics[height=.6\linewidth]{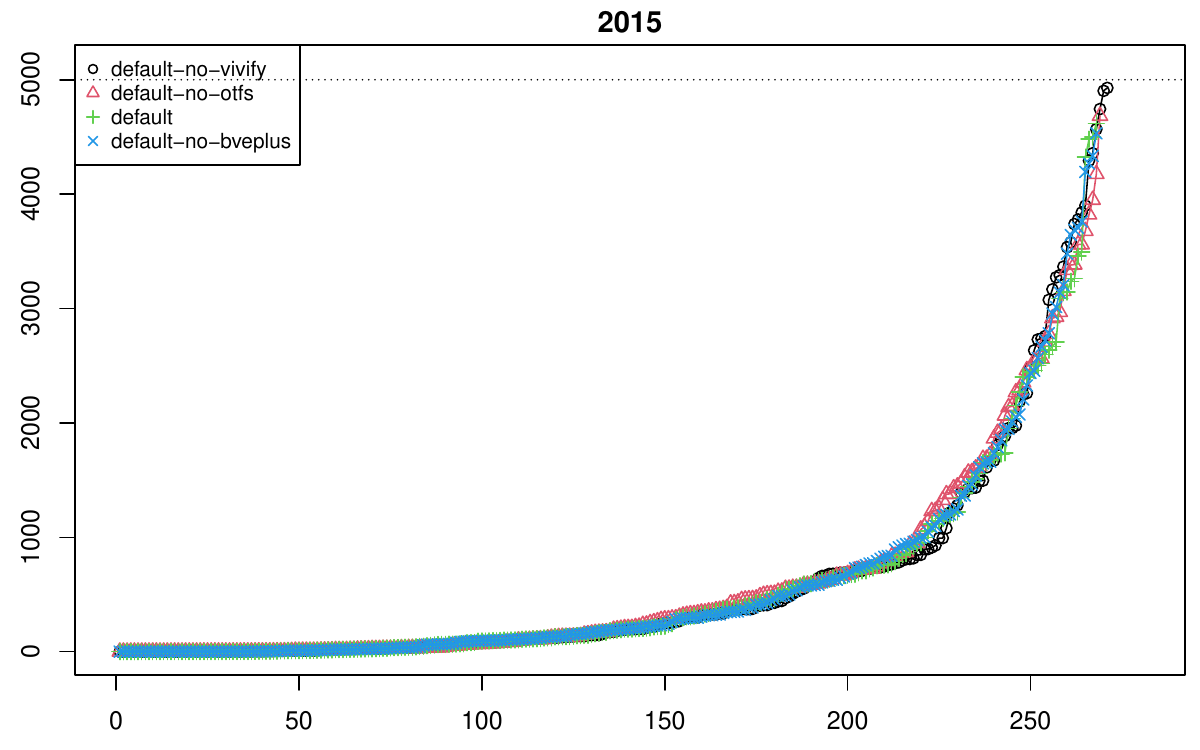}
     \end{subfigure}%
    \begin{subfigure}{.5\textwidth}
      \centering
      \includegraphics[height=.6\linewidth]{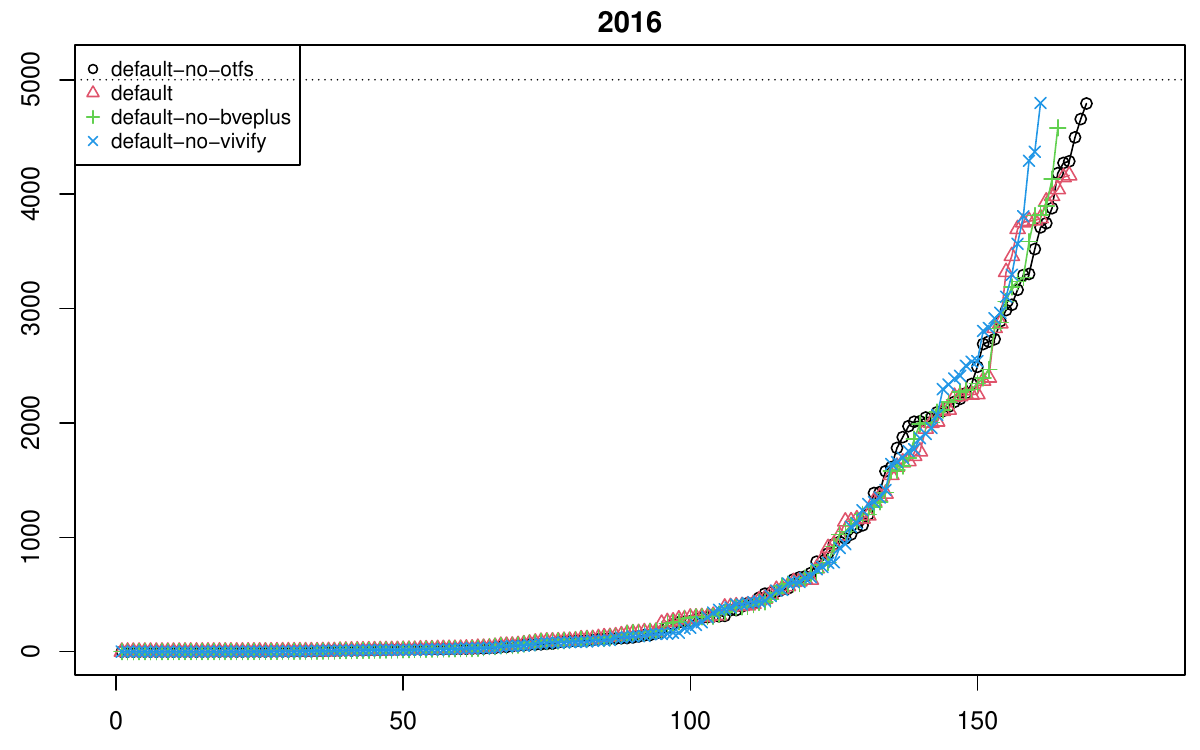}
     \end{subfigure}
     \caption{Cactus Plots of experiment ``default'' -- years 2009-2016}
    \end{figure}
  
    \begin{figure}[p]
      \centering
    \begin{subfigure}{.5\textwidth}
      \centering
      \includegraphics[height=.6\linewidth]{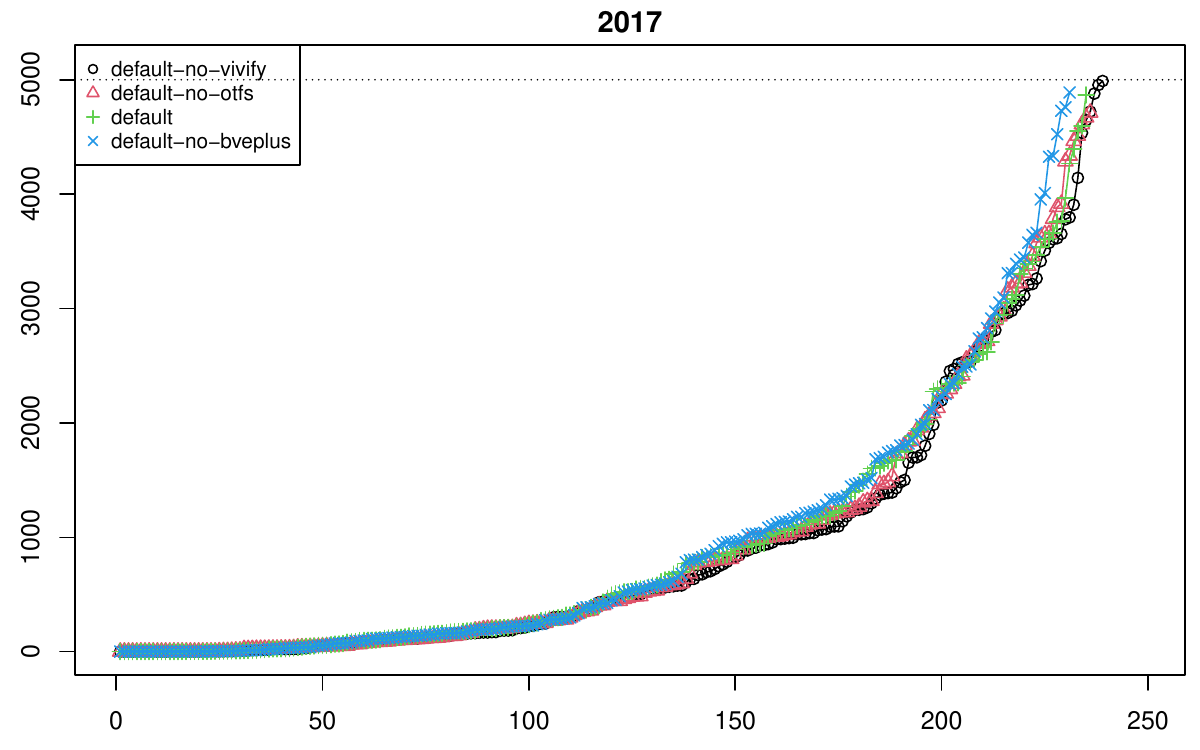}
     \end{subfigure}%
    \begin{subfigure}{.5\textwidth}
      \centering
      \includegraphics[height=.6\linewidth]{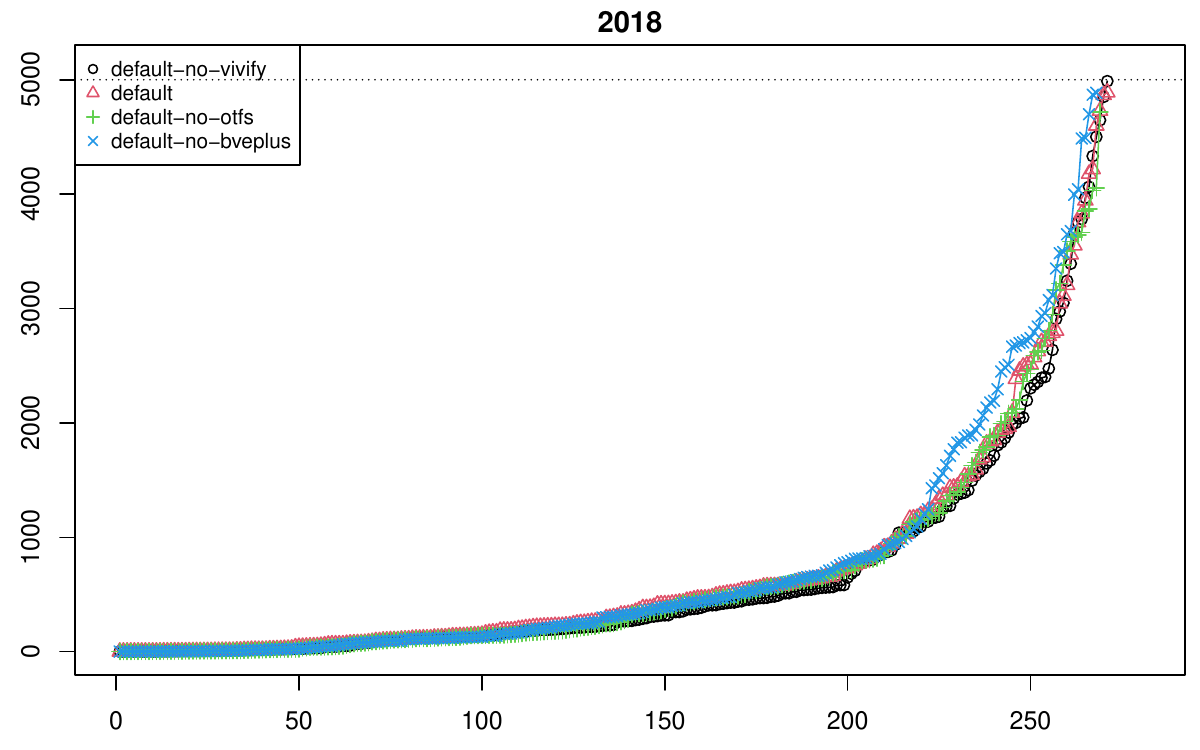}
     \end{subfigure}

    \begin{subfigure}{.5\textwidth}
      \centering
      \includegraphics[height=.6\linewidth]{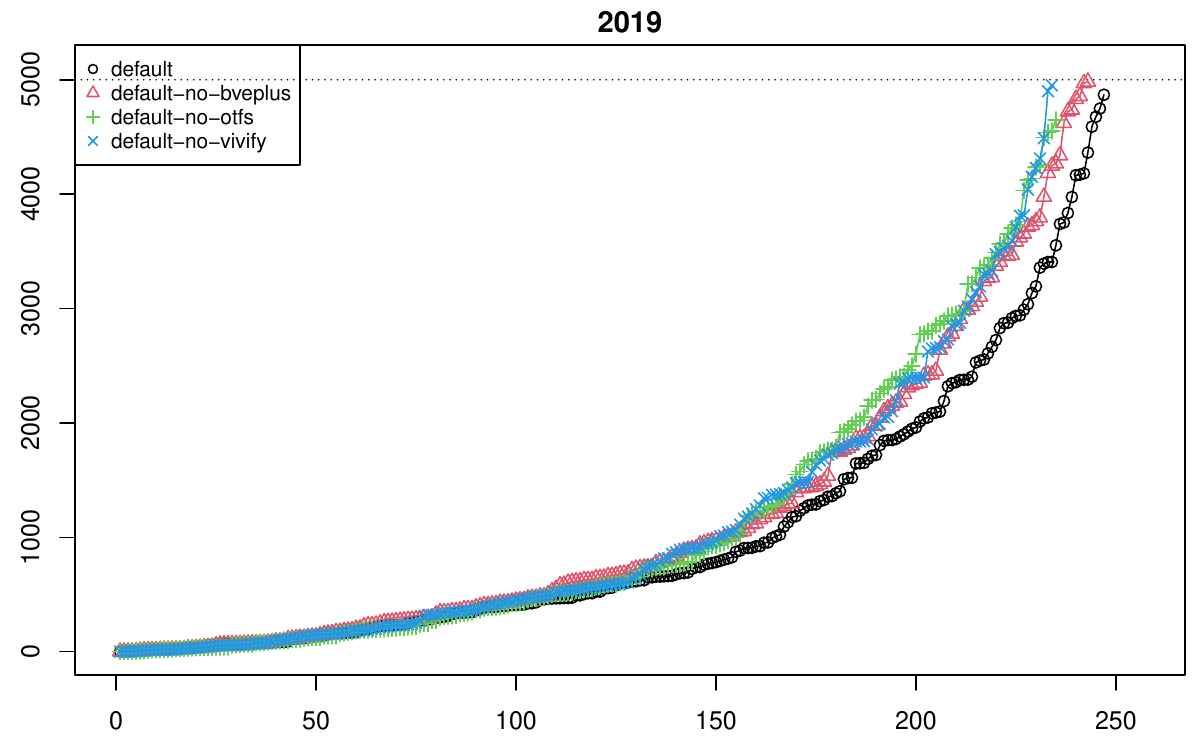}
    \end{subfigure}%
    \begin{subfigure}{.5\textwidth}
      \centering
      \includegraphics[height=.6\linewidth]{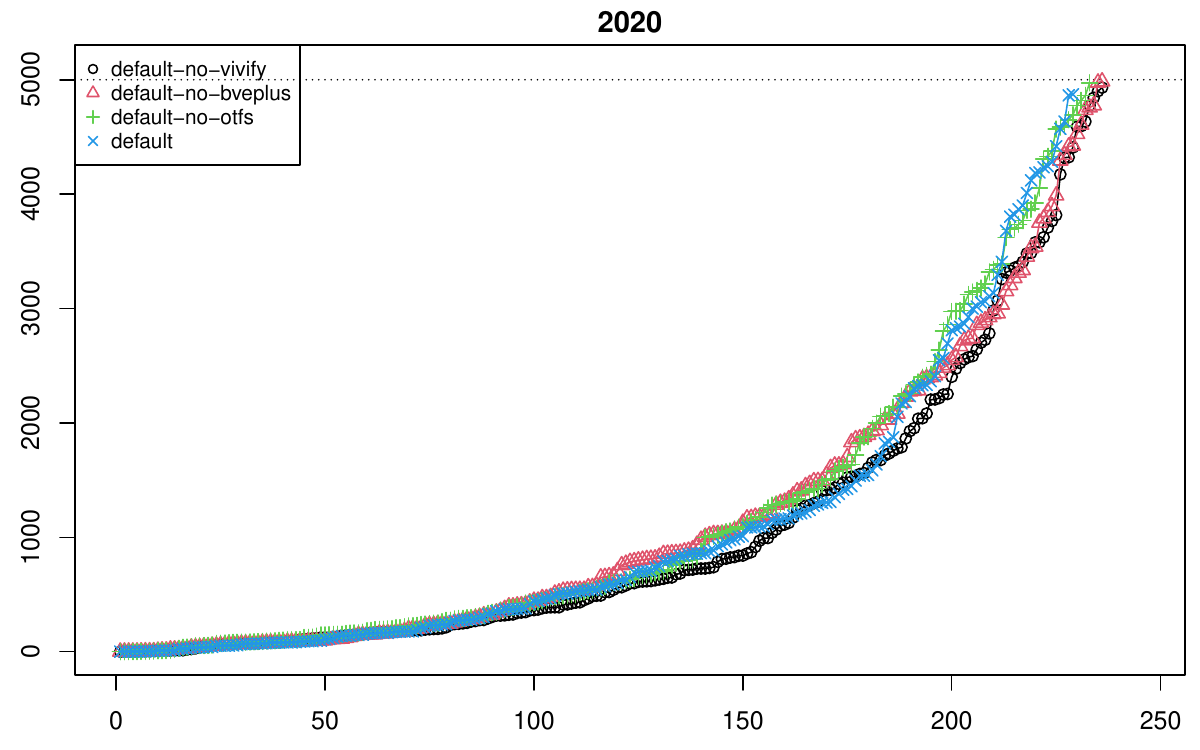}
    \end{subfigure}
    \begin{subfigure}{.5\textwidth}
      \centering
      \includegraphics[height=.6\linewidth]{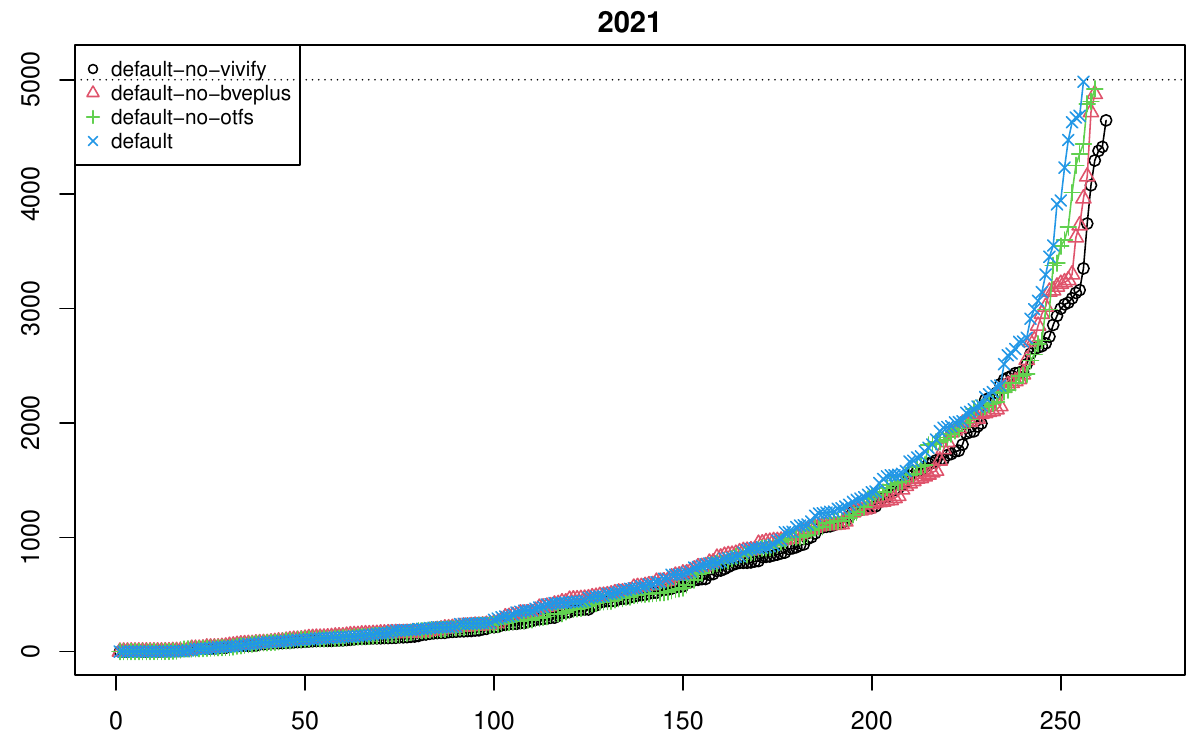}
    \end{subfigure}%
    \begin{subfigure}{.5\textwidth}
      \centering
      \includegraphics[height=.6\linewidth]{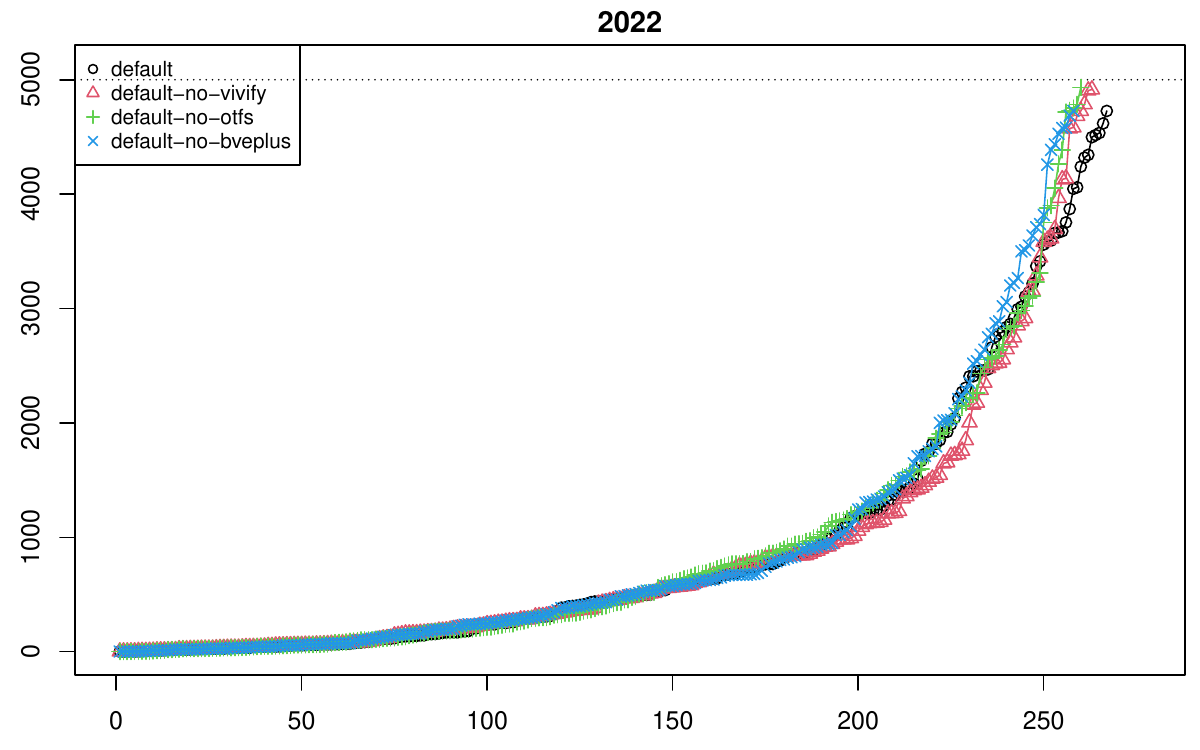}
     \end{subfigure}
    \caption{Cactus Plots of experiment ``default'' -- years 2017-2022}
    \end{figure}

\newpage
    \section{Cactus Plots - Experiment ``everything''}
    \label{app:cactus-everything}

    \begin{figure}[ht!]
      \centering
      \begin{subfigure}{.5\textwidth}
        \centering
        \includegraphics[height=.6\linewidth]{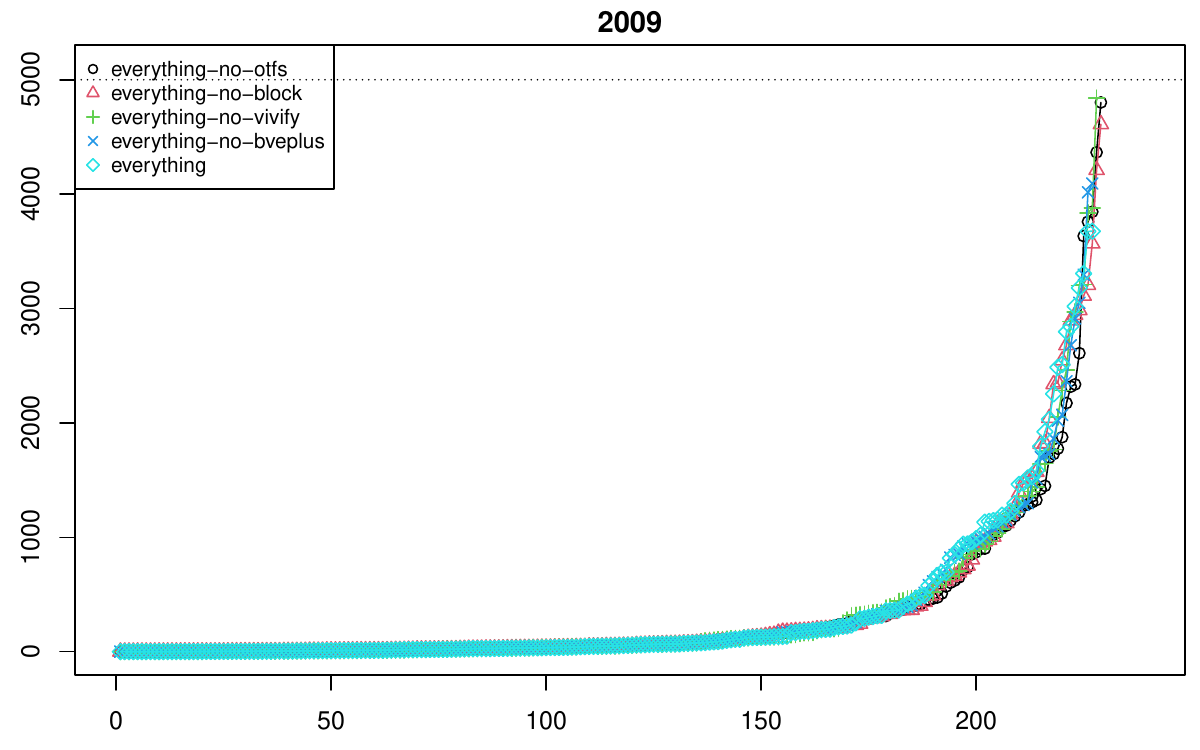}
      \end{subfigure}%
      \begin{subfigure}{.5\textwidth}
        \centering
        \includegraphics[height=.6\linewidth]{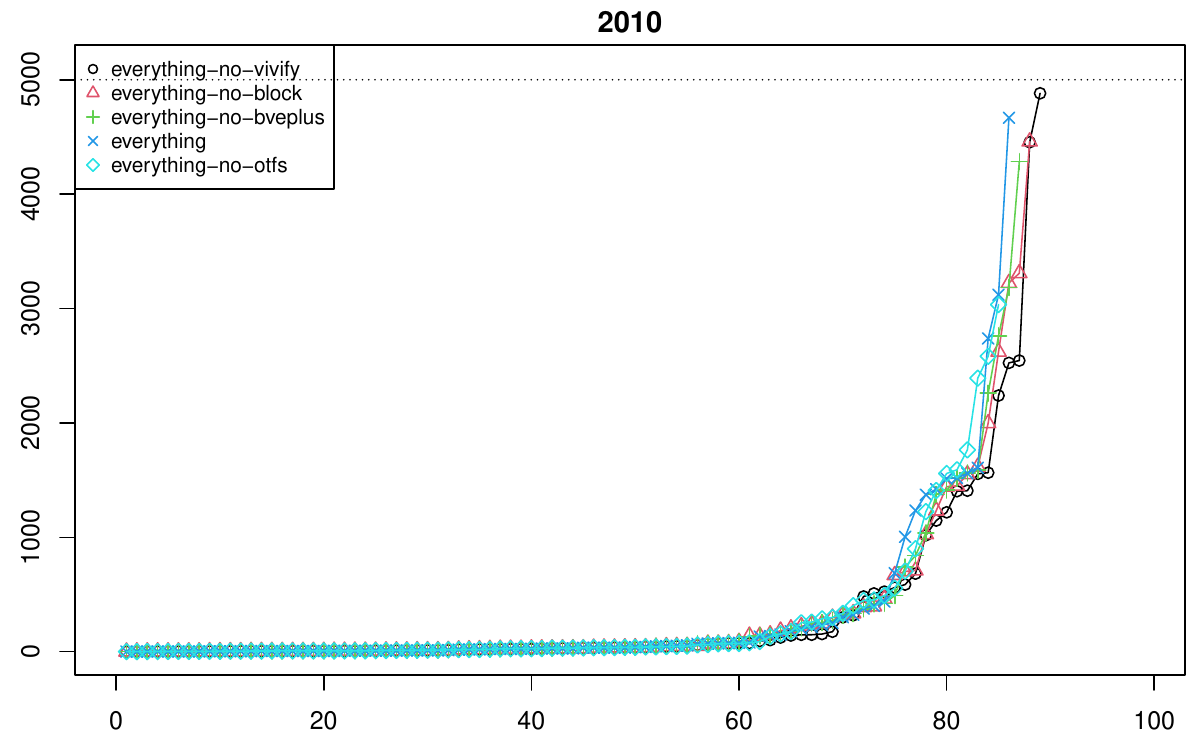}
      \end{subfigure}
      \begin{subfigure}{.5\textwidth}
        \centering
        \includegraphics[height=.6\linewidth]{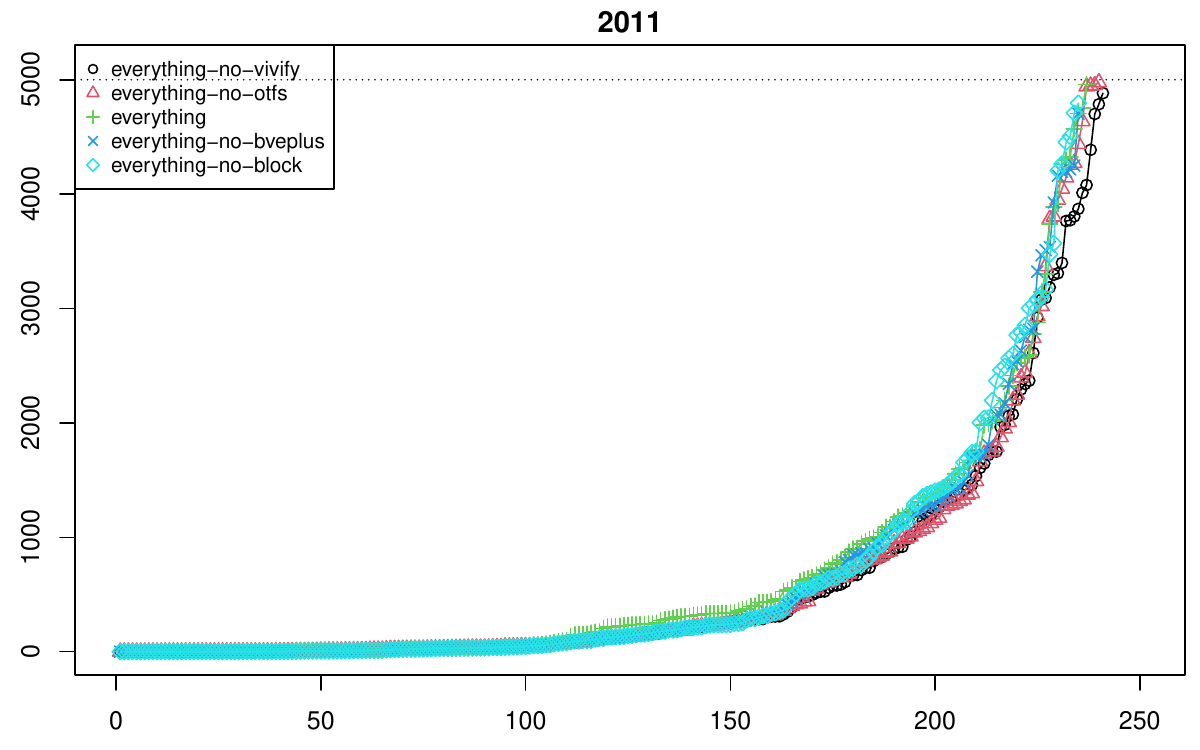}
        \end{subfigure}%
      \begin{subfigure}{.5\textwidth}
        \centering
        \includegraphics[height=.6\linewidth]{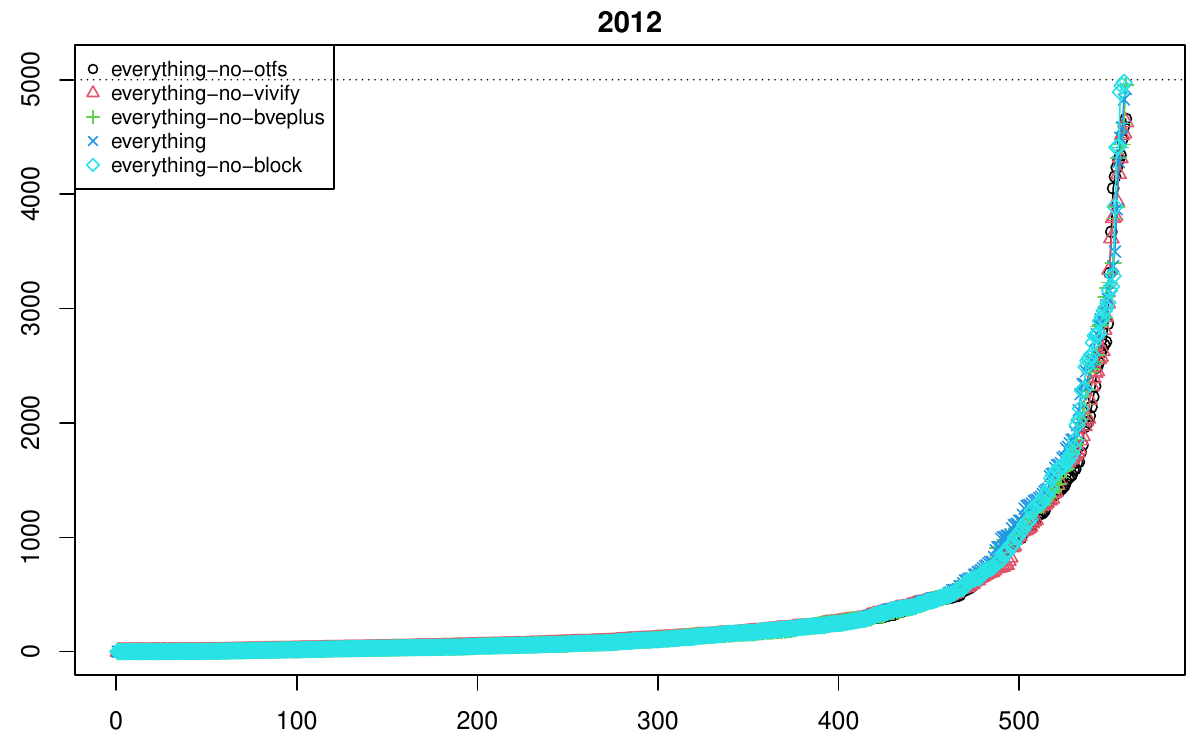}
        \end{subfigure}
      \begin{subfigure}{.5\textwidth}
        \centering
        \includegraphics[height=.6\linewidth]{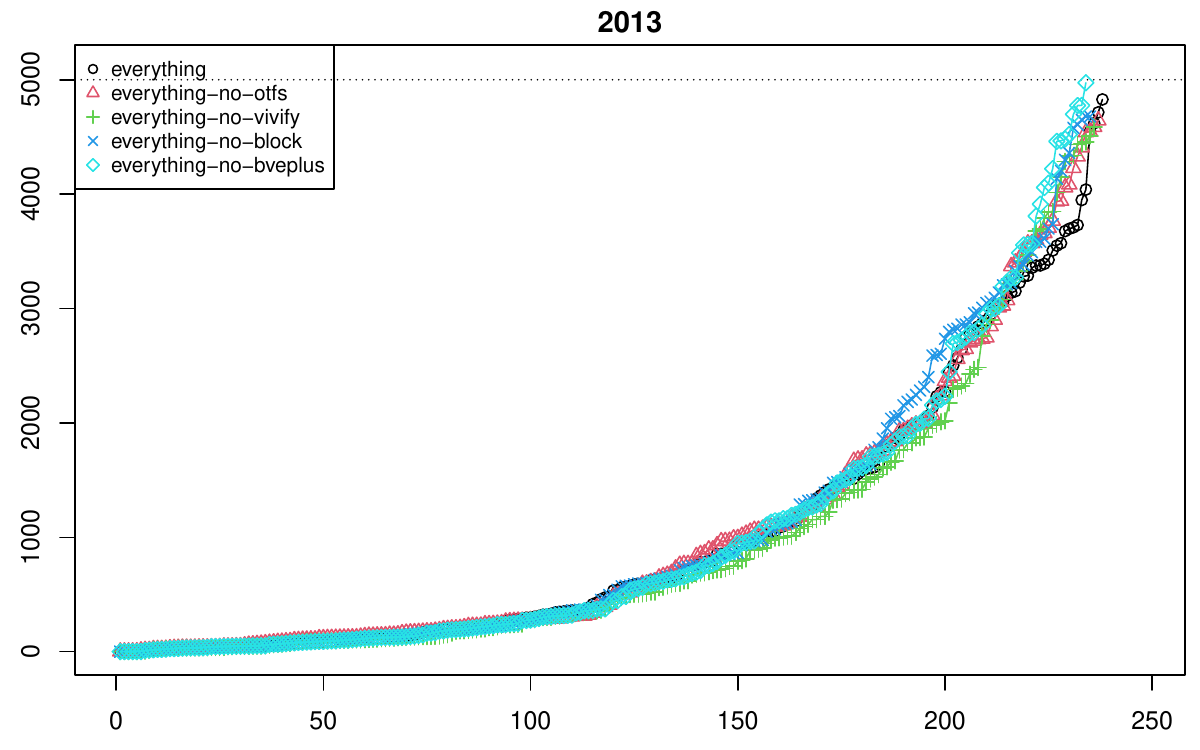}
      \end{subfigure}%
      \begin{subfigure}{.5\textwidth}
        \centering
        \includegraphics[height=.6\linewidth]{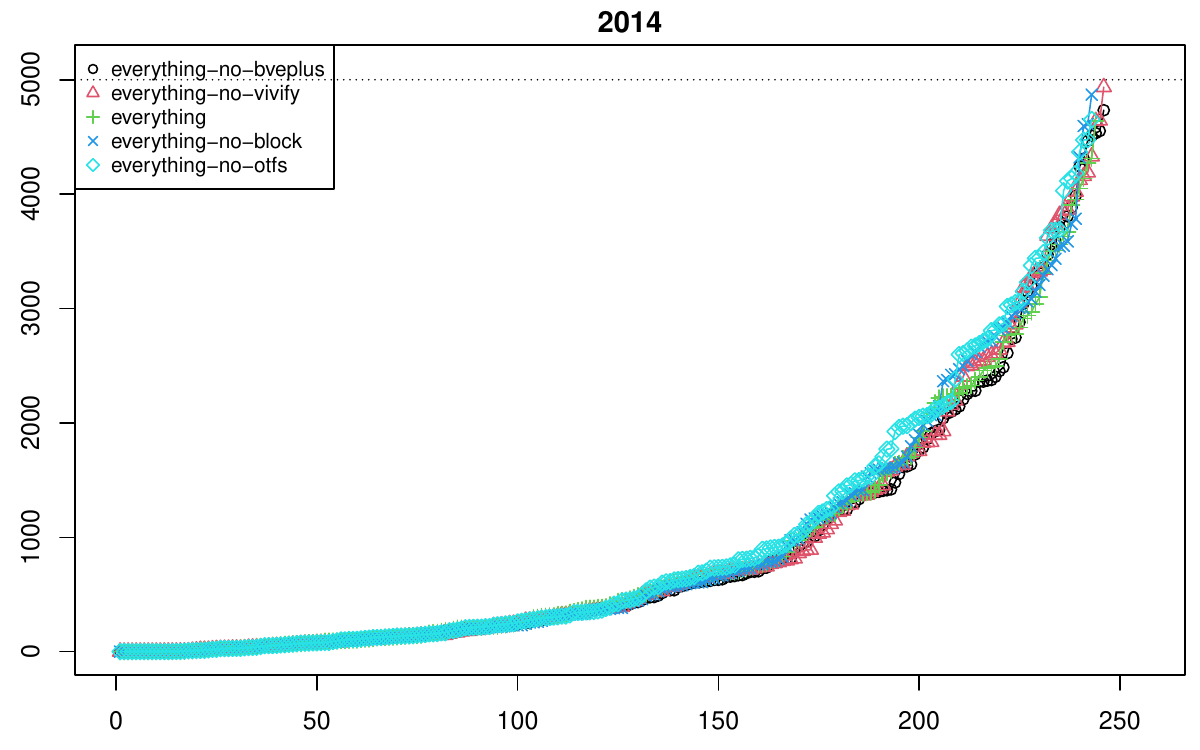}
       \end{subfigure}
      \begin{subfigure}{.5\textwidth}
        \centering
        \includegraphics[height=.6\linewidth]{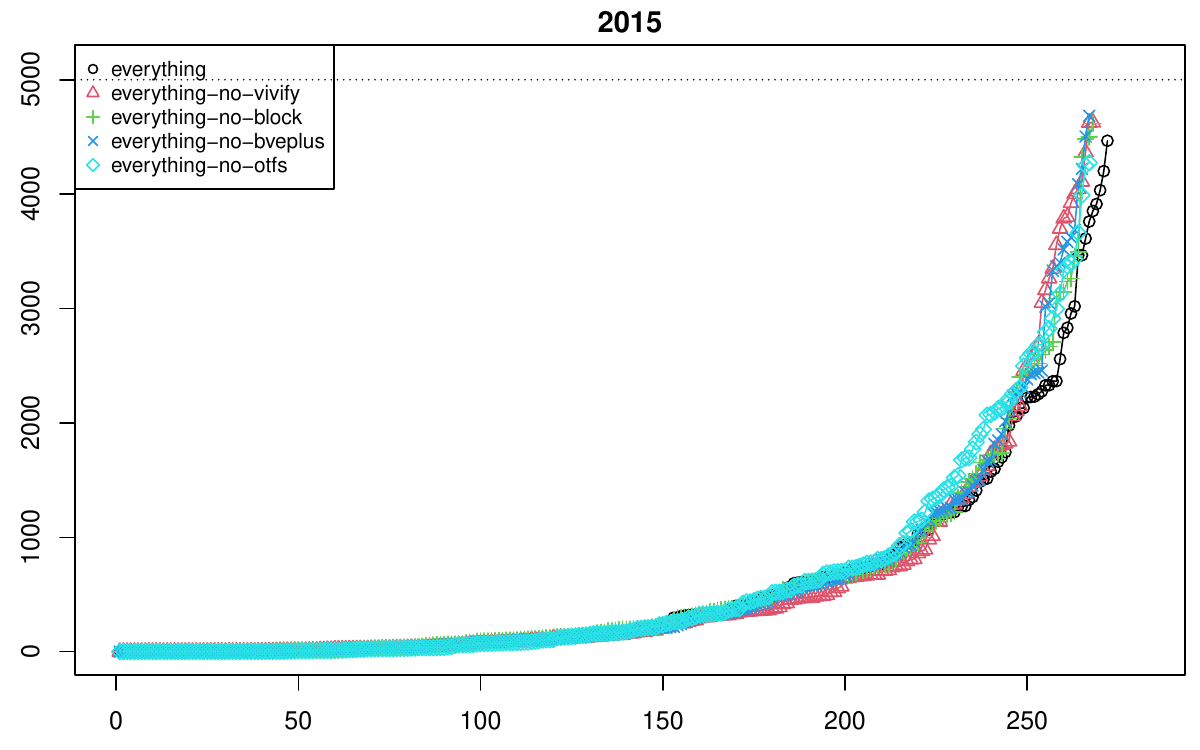}
       \end{subfigure}%
      \begin{subfigure}{.5\textwidth}
        \centering
        \includegraphics[height=.6\linewidth]{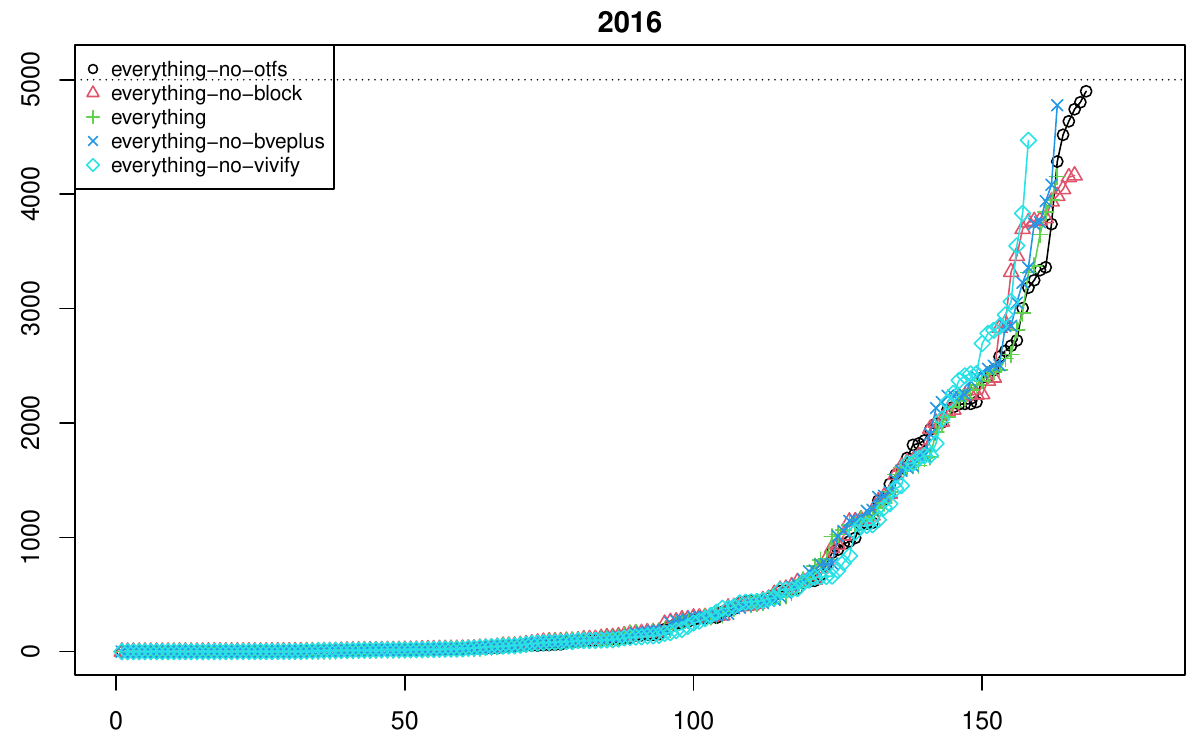}
       \end{subfigure}
       \caption{Cactus Plots of experiment ``everything'' -- years 2009-2016}
      \end{figure}

      \begin{figure}[p]
        \centering
      \begin{subfigure}{.5\textwidth}
        \centering
        \includegraphics[height=.6\linewidth]{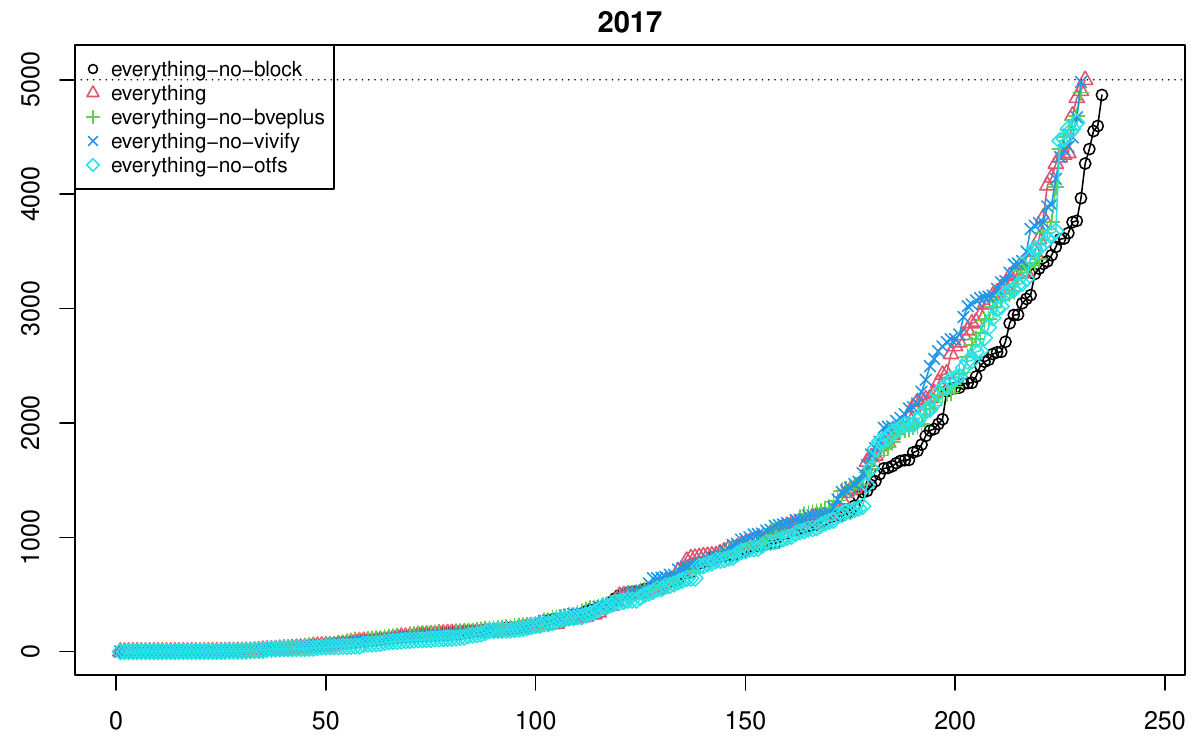}
       \end{subfigure}%
      \begin{subfigure}{.5\textwidth}
        \centering
        \includegraphics[height=.6\linewidth]{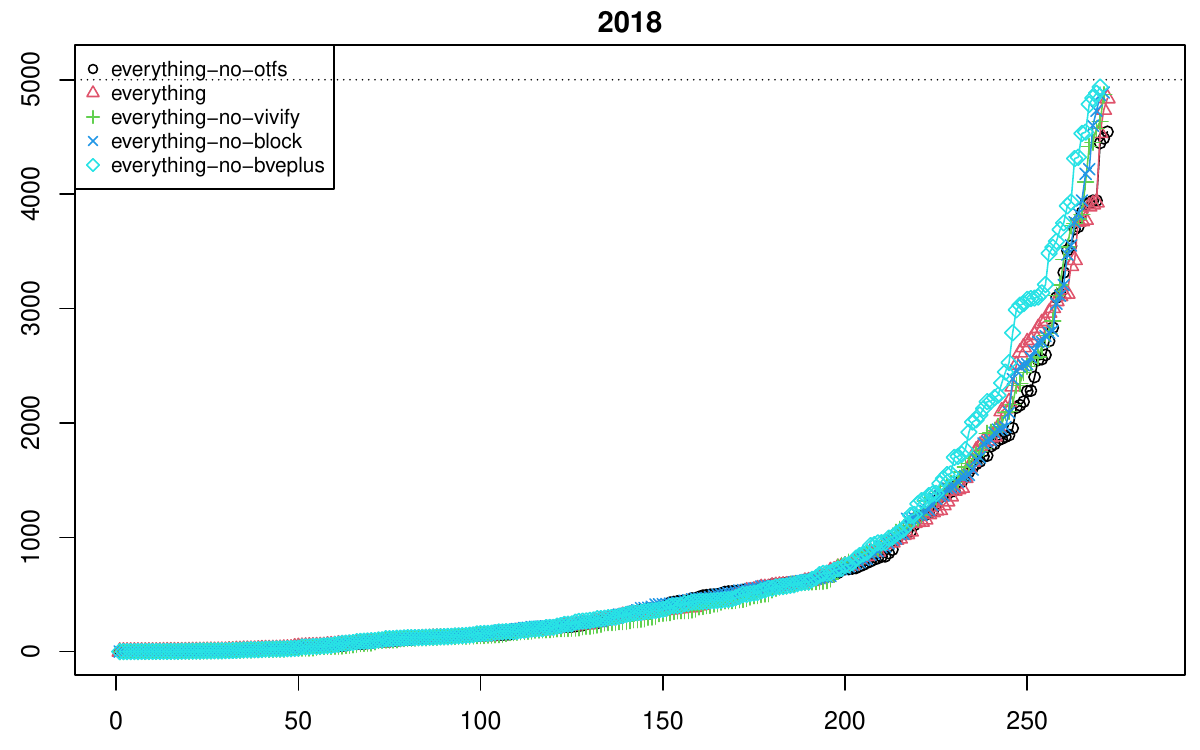}
       \end{subfigure}

      \begin{subfigure}{.5\textwidth}
        \centering
        \includegraphics[height=.6\linewidth]{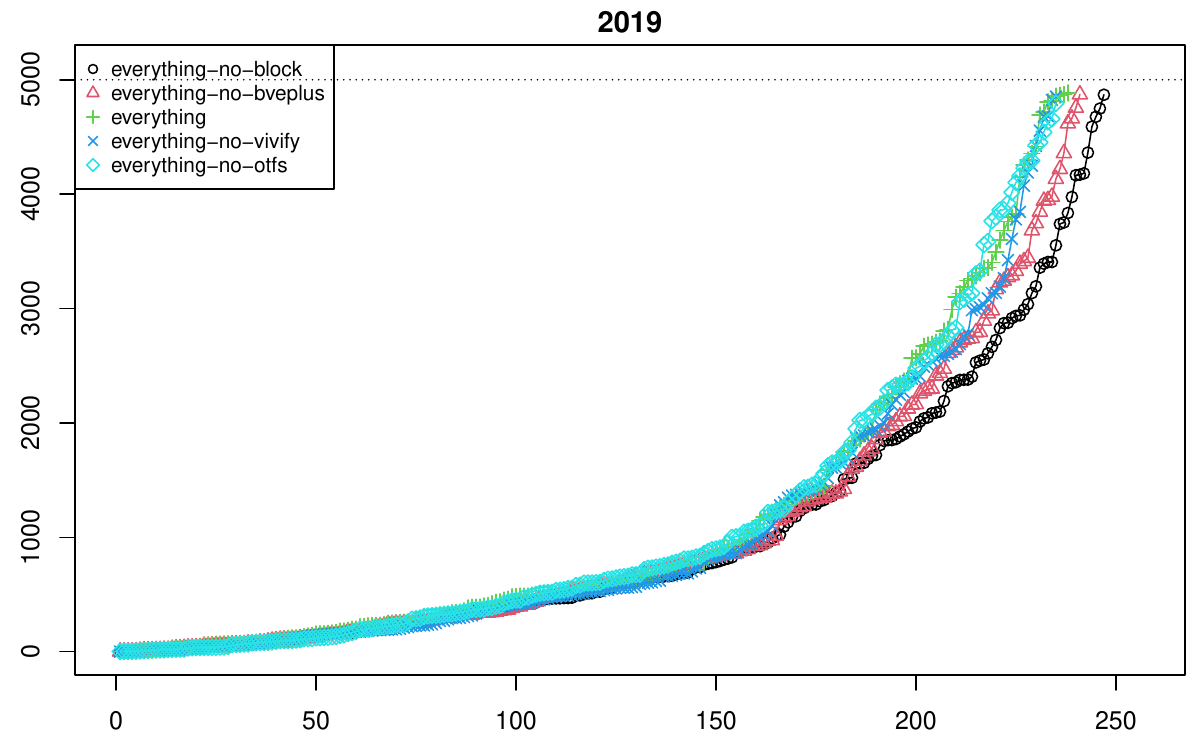}
      \end{subfigure}%
      \begin{subfigure}{.5\textwidth}
        \centering
        \includegraphics[height=.6\linewidth]{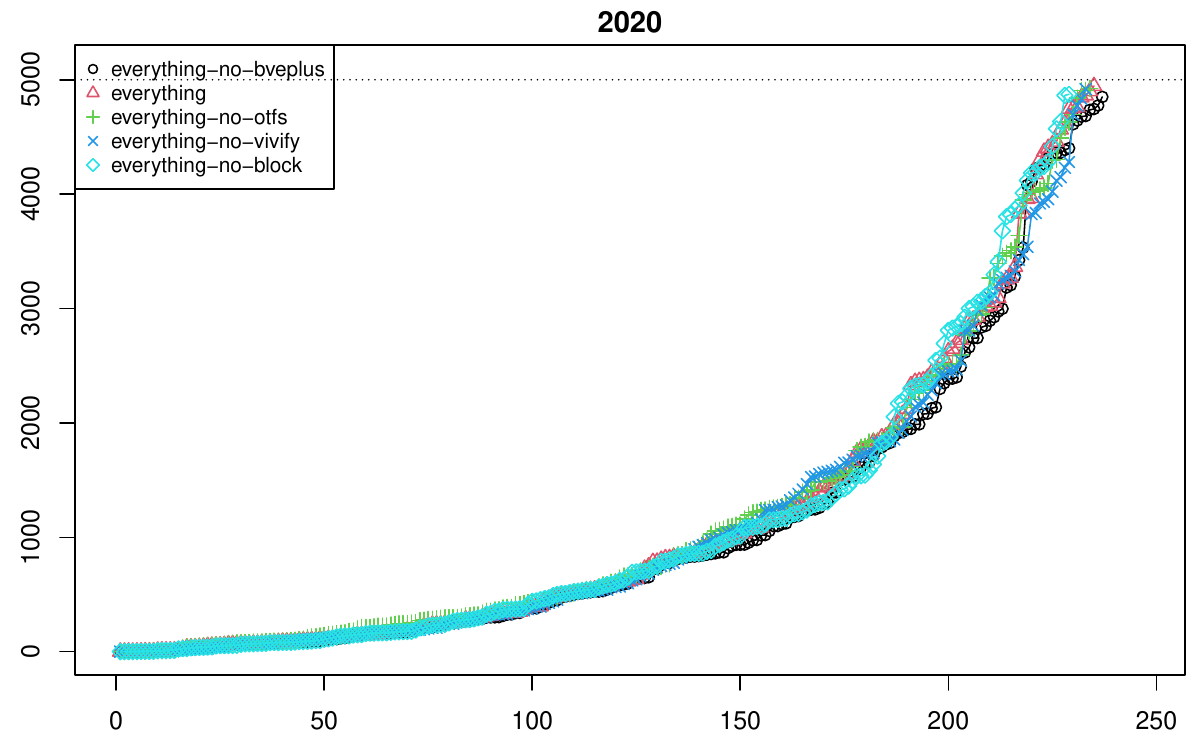}
      \end{subfigure}
      \begin{subfigure}{.5\textwidth}
        \centering
        \includegraphics[height=.6\linewidth]{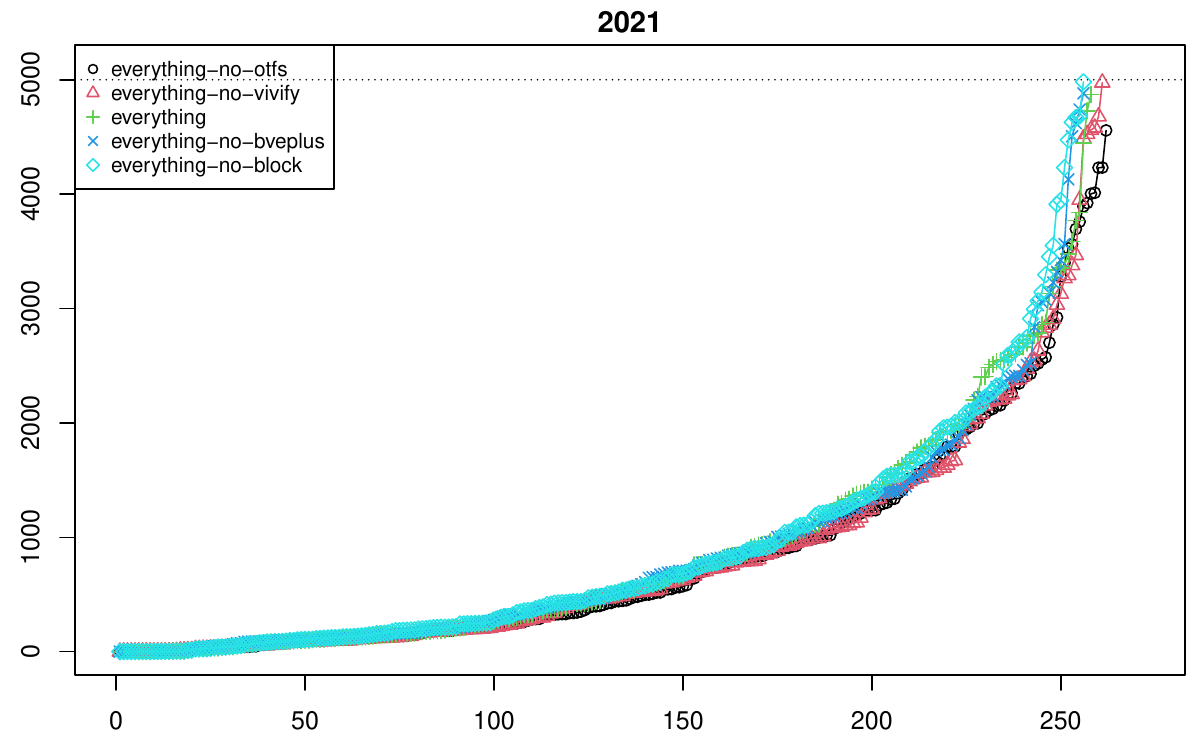}
      \end{subfigure}%
      \begin{subfigure}{.5\textwidth}
        \centering
        \includegraphics[height=.6\linewidth]{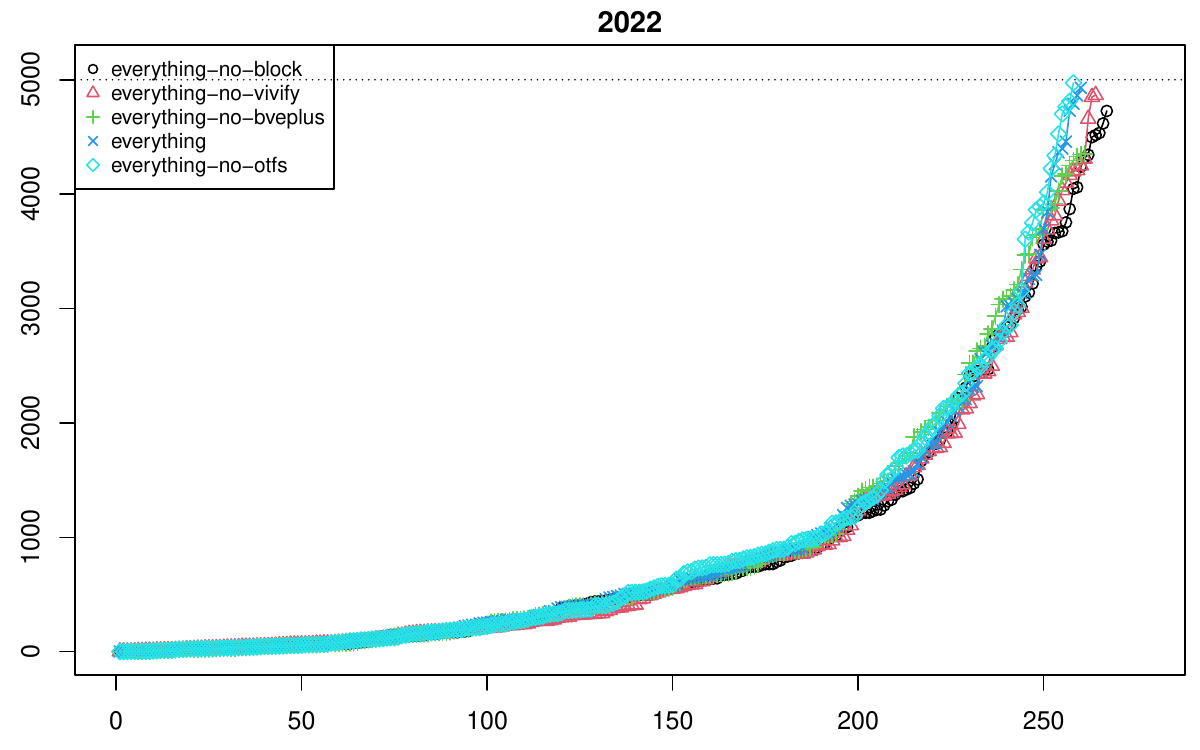}
       \end{subfigure}
      \caption{Cactus Plots of experiment ``everything'' -- years 2017-2022}
      \end{figure}

\end{document}